\documentclass[12pt]{article}
\pdfoutput=1 

\usepackage{jheppub} 
\usepackage{graphicx}  
\usepackage{dcolumn}   
\usepackage{bm}        
\usepackage{amssymb}   
\usepackage{subfigure}
\usepackage{epstopdf}
\usepackage{color}
\usepackage{slashed}
\usepackage[utf8]{inputenc}
\usepackage{multirow}
\usepackage{footnote}
\usepackage{amsmath}
\allowdisplaybreaks[4]
\usepackage{accents}
\newlength{\dhatheight}

\usepackage{hyperref}
\def\figureautorefname~#1\null{Fig.\,#1\null}
\def\tableautorefname~#1\null{Tab.\,#1\null}

\def\equationautorefname~#1\null{Eq.\,(#1)\null}

\title{Fermion Masses, Neutrino Mixing and Higgs-Mediated Flavor Violation in 3HDM with \boldmath{$S_3$} Permutation Symmetry}
\author[b]{K.S. Babu,}
\author[,a,b]{Yongcheng Wu,\footnotemark[1]}
\author[,b,c]{Shiyuan Xu\footnotemark[1]}
\footnotetext[1]{Corresponding authors}
\affiliation[a]{Department of Physics and Institute of Theoretical Physics, Nanjing Normal University, Nanjing, 210023, China}
\affiliation[b]{Department of Physics, Oklahoma State University, Stillwater, OK, 74078, USA}
\affiliation[c]{Division of Natural Sciences, Maryville College, Maryville, TN, 37804, USA}
\emailAdd{babu@okstate.edu}
\emailAdd{ycwu@njnu.edu.cn}
\emailAdd{shiyuan.xu@maryvillecollege.edu}

\abstract{
The Yukawa and scalar sectors of a general $S_3$-symmetric three-Higgs doublet model (3HDM) are investigated.
The Yukawa interactions are constructed in an $S_3$-invariant way, while the scalar potential contains $S_3$ soft-breaking terms.
Global fits to the quark/lepton masses and CKM/PMNS matrices are performed. Excellent fits to all fermion mass and mixing parameters are obtained. Both normal ordering and inverted ordering of neutrino masses are found to be admissible within the framework, with a prediction for the CP-violation phase, $\delta_{CP} \simeq 120^0$.
The fit results in the Yukawa sector are further investigated, together with the scalar sector, imposing constraints from Higgs-mediated neutral meson mixing and neutron electric dipole moment (EDM).
We explore the lowest allowed mass of the heavy Higgs bosons, consistent with these constraints, and find it to be about 17 TeV. The corresponding neutron EDM is around $1.7\times10^{-27}$ e-cm, which is within reach of proposed experiments. It is found that the constraints from the $K$-meson system dominate, while those from the $D$ meson system are marginal.
}

\begin{document}
\maketitle
\newpage

\flushbottom

\section{Introduction}
The Standard Model (SM) has withheld a variety of direct and indirect tests over decades and its success is one of the most outstanding achievements of modern physics. However, it is unlikely to be the ultimate theory since there are many questions that remain unexplained, including the existence of dark matter, nonzero neutrino masses,  the origin of matter-antimatter asymmetry, and the observed hierarchical structure in the fermion masses and mixings. These unexplained questions are strong motivations to explore new physics beyond the Standard Model (BSM).
The Higgs boson discovered at the LHC a decade ago~\cite{ATLAS:2012yve,CMS:2012qbp} provides a new window to explore possible physics beyond the SM. In the SM, the scalar sector is constructed in the simplest possible way using one scalar doublet. Extensions with extra singlet~\cite{Profumo:2007wc,Chiang:2017nmu}, doublets~\cite{Branco:2011iw,Fontes:2017zfn} as well as triplets~\cite{Georgi:1985nv,Hartling:2014zca} are widely investigated to accommodate some of the aforementioned questions.

The three-Higgs doublet model (3HDM), which is the focus of this paper, is well-motivated when viewed in the context of the flavor structure with three generations of fermions. To reduce the number of free parameters in a general 3HDM, discrete symmetries are often employed, which would also be used to explain various phenomena in flavor physics that appear to be independent. In the context of 3HDM, various authors have studied the cases with the discrete symmetry being the permutation groups $S_3$~\cite{Pakvasa:1977in,Derman:1978rx,Derman:1979nf,Wyler:1978fj,Frere:1978ds,Yahalom:1983kf,Ma:1990qh,Hall:1995es,Koide:1999mx,Lavoura:1999dn,Kubo:2003iw,Kubo:2004ps,Chen:2004rr,Kimura:2005sx,Teshima:2005bk,Koide:2005ep,Araki:2005ec,Mondragon:2007nk,Mondragon:2007af,Bhattacharyya:2010hp,Teshima:2011wg,Teshima:2012cg,Bhattacharyya:2012ze,GonzalezCanales:2012blg,GonzalezCanales:2013pdx,Das:2014fea,Das:2015sca,Emmanuel-Costa:2016vej,Kuncinas:2020wrn,Gomez-Bock:2021uyu,Khater:2021wcx,Kuncinas:2022whn}, or $S_4$~\cite{Yamanaka:1981pa,Brown:1984mq,CarcamoHernandez:2022vjk}, as well as $A_4$, the symmetry group of a regular tetrahedron~\cite{Pramanick:2017wry,Buskin:2021eig,Carrolo:2022oyg}. A complete list of realizable discrete symmetries in 3HDM has been constructed in Ref.~\cite{Ivanov:2012fp}.

In this paper, we consider the 3HDM with $S_3$ permutation symmetry, which is the simplest non-Abelian symmetry. The three generations of fermions and the three scalar doublets are assigned to the singlet and doublet representations under $S_3$ symmetry.\footnote{The three fermion families and the three scalar doublets can be naturally assigned to the triplet representations of the $S_3$. However, the $n$-dimensional representation of $S_n$ is reducible, which can be decomposed into irreducible 1-dimensional and $(n-1)$-dimensional representations. The discussions based on either triplet or singlet plus doublet are equivalent.} This is the most symmetrical assignment, as it treats all fermion families as well as the scalar doublets on equal footing. The Yukawa interactions are constructed in a $S_3$-invariant way. However, we allow for the $S_3$ symmetry to be broken softly in the scalar sector, which helps to realize the decoupling limit of scalars. We allow for CP violation both in the scalar interactions, as well as in the Yukawa couplings.  The target is to examine the scalar sector in 3HDM along with the $S_3$-symmetric Yukawa sector and confront the model against measured values of the quark masses, CKM matrix including the CP violating phase, as well as the lepton masses and PMNS mixing matrix from neutrino oscillation measurements. We introduce right-handed neutrinos to generate neutrino masses through the seesaw mechanism, which are also assigned to a singlet and doublet representations of $S_3$. The quark/lepton sector in 3HDM with $S_3$ symmetry has been studied in previous works~\cite{Teshima:2011wg,Teshima:2012cg,GonzalezCanales:2013pdx,Das:2015sca}, which, however, focused on simplified cases by introducing additional symmetries in different sectors or by assuming CP to be conserved in the scalar sector. Owing to these simplifications, the Yukawa interactions and the Higgs potentials of those analyses are not the most general ones.  Here we construct the most general $S_3$-invariant interactions and analyze them in detail.

In the current study, we analyze the quark and lepton sectors simultaneously, taking into account the constraints on the heavy Higgs boson massed from neutral meson mixings. With the CP phases that arise naturally in the vacuum expectation values (VEVs) of the Higgs fields, which are needed to reproduce CKM CP violation, we find that there are important new contributions to the neutron electric dipole moment (EDM). Our successful fit of the fermion mass and mixing parameters show that this model can provide an understanding of the fermion mass hierarchy and the origin of CP violation, and it predicts the value of the CP phase ($\delta_{CP}$) appearing in the PMNS matrix.  One issue that we address here carefully is the lowest allowed value of the heavy Higgs boson mass, consistent with flavor changing neutral currents and the neutron EDM.  We find that the lightest value the heavy Higgs boson could have, consistent with these constraints, is about 17 TeV. The dominant constraint comes from $K^0-\overline{K}^0$ mixing, while those from the $D^0$ and $B_{d,s}^0$ meson systems are marginal. The corresponding neutron EDM is around $1.7\times10^{-27} e\cdot{\rm cm}$, which is about one order of magnitude below the current experimental limit.

The rest of the paper is organized as follows. In~\autoref{sec:model}, we discuss the model setup including the scalar sector and the Yukawa sector. Theoretical constraints including bounded from below (BFB) conditions and perturbative unitarity are also discussed here. The constraints from experimental measurements including neutral meson mixing and neutron EDM are discussed in~\autoref{sec:fcnc-edm}. With all the constraints we considered, the best fit result in quark and lepton sections as well as the scalar sector are listed in~\autoref{sec:Results} where we derive the lowest heavy Higgs mass to be about 17 TeV. We conclude in~\autoref{sec:conclusion}. Some tedious formulae for the minimization conditions of the Higgs potential, scalar mass matrices, and numerical values of various Yukawa coupling matrices are listed in several appendices.

\section{The Model Setup}
\label{sec:model}
\subsection{The Scalar Sector}
We denote three Higgs doublets of the
3HDM model as $\Phi_i$, $i=1,2,3$. Without loss of generality, we assume that $(\Phi_1,\Phi_2)$ forms a doublet of the $S_3$ symmetry, and $\Phi_3$ is a singlet. Then the corresponding scalar potential which is invariant under the $SU(3)_c \times SU(2)_L \times U(1)_Y \times S_3$ is~\cite{Kuncinas:2020wrn,Kubo:2004ps,Emmanuel-Costa:2017zkm}
\begin{align}
V_0= & \mu_0^2(\Phi_1^\dagger \Phi_1+\Phi_2^\dagger \Phi_2)+\mu_1^2(\Phi_3^\dagger \Phi_3) +\lambda_1(\Phi_1^\dagger \Phi_1+\Phi_2^\dagger \Phi_2)^2 \nonumber\\
&+\lambda_2(\Phi_1^\dagger \Phi_2-\Phi_2^\dagger \Phi_1)^2 +\lambda_3[(\Phi_1^\dagger \Phi_1-\Phi_2^\dagger \Phi_2)^2+(\Phi_1^\dagger \Phi_2+\Phi_2^\dagger \Phi_1)^2] \nonumber\\
& +[\lambda_4e^{i\beta_4}[(\Phi_3^\dagger \Phi_1)(\Phi_1^\dagger \Phi_2+\Phi_2^\dagger \Phi_1)+(\Phi_3^\dagger \Phi_2)(\Phi_1^\dagger \Phi_1-\Phi_2^\dagger \Phi_2)]+h.c.] \nonumber\\
& +\lambda_5[(\Phi_3^\dagger \Phi_3)(\Phi_1^\dagger \Phi_1+\Phi_2^\dagger \Phi_2)]+\lambda_6[(\Phi_3^\dagger \Phi_1)(\Phi_1^\dagger \Phi_3)+(\Phi_3^\dagger \Phi_2)(\Phi_2^\dagger \Phi_3)]\nonumber \\
& +[\lambda_7e^{i\beta_7}[(\Phi_3^\dagger \Phi_1)^2+(\Phi_3^\dagger \Phi_2)^2]+h.c.]+\lambda_8(\Phi_3^\dagger \Phi_3)^2.
\label{equ:potential_S3}
\end{align}
The $S_3$ symmetry is allowed to be softly broken by dimension-two operators~\cite{Kuncinas:2020wrn,Kubo:2004ps}:
\begin{align}
V_{\rm soft}= & \mu_2^2(\Phi_1^\dagger \Phi_1-\Phi_2^\dagger \Phi_2)+\frac{1}{2}(\mu_3^2e^{i\alpha_{3}}\Phi_1^\dagger \Phi_2+h.c.)+\frac{1}{2}(\mu_4^2e^{i\alpha_4}\Phi_1^\dagger \Phi_3+h.c.)\nonumber\\
& +\frac{1}{2}(\mu_5^2e^{i\alpha_5}\Phi_2^\dagger \Phi_3+h.c.).
\label{equ:potential_soft}
\end{align}
Hence, the total scalar potential $V$ is the sum of the $S_3$ invariant part and the soft-breaking part:
\begin{align}
    V=V_0+V_{\rm soft}.
\end{align}
The real parameters in the potential are thus
\begin{align}
\text{Couplings:}\qquad    &\mu_{0,1,\cdots,5},\lambda_{1,2,\cdots,8},\nonumber\\
\text{Phases:}\qquad    &\alpha_{3,4,5},\beta_{4,7}.
\end{align}

The minimization conditions for the above scalar potential are discussed in Ref.~\cite{Kuncinas:2020wrn}. We parametrize the three Higgs doublets, after spontaneous symmetry breaking, as
\begin{align}
\Phi_1 = \left(\begin{array}{c}
    \omega_1^+\\
    \frac{v_1e^{i\theta_1} + \phi_1 + i\xi_1}{\sqrt{2}}
\end{array}\right),\quad
\Phi_2 = \left(\begin{array}{c}
    \omega_2^+\\
    \frac{v_2e^{i\theta_2} + \phi_2 + i\xi_2}{\sqrt{2}}
\end{array}\right),\quad
\Phi_3 = \left(\begin{array}{c}
    \omega_3^+\\
    \frac{v_3+\phi_3+i\xi_3}{\sqrt{2}}
\end{array}\right),
\label{equ:scalar_vevs}
\end{align}
where $v_{1,2,3}$ and $\theta_{1,2}$ are real parameters and where $\sqrt{v_1^2+v_2^2+v_3^2}=246\ \mathrm{GeV}$. Note that a possible phase in $v_3$ has been removed by a gauge rotation.
Including the CP violating phases  in the potential in~\autoref{equ:potential_S3} and~\autoref{equ:potential_soft} and $\theta_{1,2}$ in the VEVs in~\autoref{equ:scalar_vevs},
we will have five constraint equations from minimization:
\begin{align}
    \frac{\partial V}{\partial v_i} &= 0, \quad i=1,2,3,\nonumber\\
    \frac{\partial V}{\partial \theta_j} &= 0, \quad j=1,2.\label{min equation}
\end{align}
These five minimization conditions corresponding to the three VEVs and two phase factors are used to eliminate $\{\mu_0, \mu_1, \mu_2, \mu_3, \mu_4\}$. The detailed expression for $\mu_{0,1,2,3,4}$ in terms of other parameters are listed in~\autoref{mini1}-\autoref{mini5}. With these conditions, the free parameters we have in the scalar sector are
\begin{align}
    \text{VEVs:}&\qquad v_1,\,v_2,\,v_3,\ \text{with }\sqrt{v_1^2+v_2^2+v_3^2}=246\,{\rm GeV},\nonumber \\
    \text{Couplings:}&\qquad \mu_5,\,\lambda_{1,\cdots,8},\nonumber \\
    \text{Phases:}&\qquad \alpha_{3,4,5},\,\beta_{4,7},\,\theta_{1,2}.
\end{align}
Hence, we are left with the one dimensionful parameter $\mu_5$, which can be taken to large values to realize the decoupling limit of scalars.

\subsection*{Scalar Masses in the Higgs Basis\label{section 3.3}}
It is convenient for us to first transform the $\Phi_I$ fields into the Higgs basis in which only one doublet contains all the VEV and the Nambu-Goldstone boson.  This field, although not a mass eigenstate, is very close to being the scalar doublet in the SM. The transformation from the $S_3$ basis $(\Phi_1,\Phi_2,\Phi_3)$ to the Higgs basis $(H_1,H_2,H_3)$ is defined as\footnote{Note that the transformation is not uniquely defined. One has the freedom to rotate again in $H_2$-$H_3$ plane.}
\begin{align}
\begin{pmatrix}
H_1\\
H_2\\
H_3
\end{pmatrix}=R_H\begin{pmatrix}
    \Phi_1\\
    \Phi_2\\
    \Phi_3\\
\end{pmatrix},\quad R_H = \begin{pmatrix}
    \frac{e^{-i\theta_1}v_1}{v} & \frac{e^{-i\theta_2}v_2}{v} & \frac{v_3}{v} \\
    0 & \frac{e^{-i\theta_2}v_3}{v_{23}} &  -\frac{v_2}{v_{23}} \\
    -\frac{e^{-i\theta_1}v_{23}}{v}  & \frac{e^{-i\theta_2}v_{1}v_2}{vv_{23}}  & \frac{v_1v_3}{vv_{23}}
    \end{pmatrix}.
\label{equ:scalar_higgs_basis}
\end{align}
with $v_{23}\equiv\sqrt{v_2^2+v_3^2}, \,v_{12}\equiv\sqrt{v_1^2+v_2^2}, \,v\equiv\sqrt{v_1^2+v_2^2+v_3^2}$.
Here the $H_1,H_2,H_3$ fields are the doublets in the Higgs basis denoted as
\begin{align}
H_1=\begin{pmatrix}
\rho_1^+ \\
(v+\eta_1+i\chi_1) / \sqrt{2}
\end{pmatrix},  \quad H_a = \begin{pmatrix}
    \rho_a^+ \\
    (\eta_a+i\chi_a)/\sqrt{2}
\end{pmatrix},\quad a=2,3.\label{equ:Higgs_doublet_Higgs_basis}
\end{align}
Note that the unitary matrix $R_H$ can also block-diagonalize the mass matrix of charged scalars as well as the neutral CP-odd bosons (in the CP-conserving case). However, usually, one needs additional rotations to diagonalize the CP-even scalar mass matrix.

Focusing on the CP-violating case, the CP-even scalars $\eta_i$ will mix with the CP-odd ones $\chi_j$. The mass matrix for these neutral scalars in basis of $(\eta_1,\eta_2,\eta_3,\chi_1,\chi_2,\chi_3)$ is
\begin{align}
    M^2 = \begin{pmatrix}
        M^2_{11} & M^2_{12} & M^2_{13} & 0 & M^2_{15} & M^2_{16} \\
        M^2_{21} & M^2_{22} & M^2_{23} & 0 & M^2_{25} & M^2_{26} \\
        M^2_{31} & M^2_{32} & M^2_{33} & 0 & M^2_{35} & M^2_{36} \\
        0 & 0 & 0 & 0 & 0 & 0 \\
        M^2_{51} & M^2_{52} & M^2_{53} & 0 & M^2_{55} & M^2_{56} \\
        M^2_{61} & M^2_{62} & M^2_{63} & 0 & M^2_{65} & M^2_{66} \\
    \end{pmatrix}~.
    \label{eq:neutral}
\end{align}
The detailed expression for each element of $M^2$ can be found in~\autoref{D1}-\autoref{D15}. Note that the fourth row and column are zero as the corresponding neutral scalar is the Goldstone boson absorbed by the $Z^0$ gauge boson. The neutral boson mass matrix of Eq. (\ref{eq:neutral}) is assumed to be diagonalized by an orthogonal matrix $R_0$ as
\begin{align}
    R_0^T M^2 R_0 = {\rm diag}(m_1^2,\cdots,m_5^2,0),
\label{equ:scalar_neutral_mass_rotation}
\end{align}
where we organize the massless Goldstone ($G^0$) field as the last entry:
\begin{align}
    \label{equ:neutral_scalar_rotation}
    \begin{pmatrix}
        \eta_1\\
        \eta_2\\
        \eta_3\\
        \chi_1\\
        \chi_2\\
        \chi_3
    \end{pmatrix} = R_0\cdot \begin{pmatrix}
        h_1\\
        h_2\\
        h_3\\
        h_4\\
        h_5\\
        G^0
    \end{pmatrix}.
\end{align}
with the $h_i$ fields being the mass eigenstates. Here, without loss of generality, we also identify the $h_1$ as the SM-like Higgs boson, with all the couplings being very close to that of the SM Higgs.

Separately, the mass matrix $M_{\pm}^2$ of the charged scalars in the Higgs basis $(\rho_1^\pm,\rho_2^\pm,\rho_3^\pm)$ reads
\begin{align}
    M_\pm^2 = \begin{pmatrix}
        0 & 0 & 0 \\
        0 & M_{\pm 22}^2 & M_{\pm 23}^2 \\
        0 & M_{\pm 32}^2 & M_{\pm 33}^2
    \end{pmatrix}.
    \label{eq:charged}
\end{align}
The detailed expressions for the elements of $M_\pm^2$ can be found in~\autoref{charged mass 16}-\autoref{charged mass 18}. Here the first row and column are zero as this corresponds to the charged Goldstons eaten by the $W^\pm$ gauge bosons. The mass matrix for the charged scalars of Eq. (\ref{eq:charged}) is diagonalized by a unitary matrix $R_\pm$ as
\begin{align}
    R_\pm^\dagger M_\pm^2 R_\pm = {\rm diag}(m_{\pm,1}^2,m_{\pm,2}^2,0),
\label{equ:scalar_charge_mass_rotation}
\end{align}
corresponding to the transformation
\begin{align}
    \begin{pmatrix}
        \rho_1^+\\
        \rho_2^+\\
        \rho_3^+
    \end{pmatrix} = R_\pm\cdot \begin{pmatrix}
    h_1^+\\
    h_2^+\\
    G^+
\end{pmatrix}.
\end{align}

\subsection{The Yukawa Sector}
\label{sec:yukawa_lagrangian}
To write down the Yukawa couplings of the model, we denote the quark and lepton fields, all with 3 generations, as
\begin{align}
Q_L^i=\left(\begin{array}{c}
    u^i\\
    d^i
\end{array}\right)_L,\ u_R^i,\ d_R^i,\ L_L^i=\left(\begin{array}{c}
    \nu^i\\
    \ell^i
\end{array}\right)_L,\ \ell_R^i,\ \nu_R^i,
\label{equ:fermion_S3}
\end{align}
with $i=1,2,3$. Here $\nu_R^i$ are the right-handed neutrinos introduced to realize the seesaw mechanism for small neutrino masses. The three generations of the above fields decompose into a doublet (for $i=1,2$) and a singlet (for $i=3$) of $S_3$ group, similar to that of the Higgs fields $\Phi_i$.
The most general Yukawa interactions that preserve $S_3$ symmetry are thus~\cite{Kubo:2003iw}:
\begin{align}
\mathcal{L}_Y = \mathcal{L}_{Y_d} + \mathcal{L}_{Y_u} + \mathcal{L}_{Y_\ell} + \mathcal{L}_{Y_\nu}
\end{align}
with
\begin{subequations}
\begin{align}
\mathcal{L}_{Y_{d}}=&-Y_1^d\overline{Q}^I_L \Phi_3 d^I_{R} - Y_2^d\left(\overline{Q}_L^I\sigma^1_{IJ}\Phi_1 d_{R}^J + \overline{Q}^I_L\sigma^3_{IJ}\Phi_2d_{R}^J\right)\nonumber \\
&-Y_3^d\overline{Q}^3_L\Phi_3d_{R}^3-Y_4^d\overline{Q}^3_L\Phi_Id_{R}^I - Y_5^d\overline{Q}^I_L \Phi_Id_{R}^3 + h.c.,\\ \label{d Yukawa Lag}
\mathcal{L}_{Y_{u}}=&-Y_1^u\overline{Q}^I_L\tilde{\Phi}_3u_{R}^I - Y_2^u\left(\overline{Q}^I_L\sigma^1_{IJ}\tilde{\Phi}_1u_{R}^J +\overline{Q}^I_L\sigma^3_{IJ}\tilde{\Phi}_2u_{R}^J\right)\nonumber \\
&-Y_3^u\overline{Q}^3_L\tilde{\Phi}_3u_{R}^3 - Y_4^u\overline{Q}^3_L\tilde{\Phi}_I u_{R}^I - Y_5^u\overline{Q}^I_L\tilde{\Phi}_Iu_{R}^3 + h.c.,\\ \label{u Yukawa Lag}
\mathcal{L}_{Y_{\ell}}=&-Y_1^\ell\overline{L}^I_L \Phi_3 \ell_{R}^I - Y_2^\ell\left(\overline{L}^I_L\sigma^1_{IJ}\Phi_1\ell_{R}^J+\overline{L}^I_L\sigma^3_{IJ}\Phi_2\ell_{R}^J\right)\nonumber \\
&-Y_3^\ell\overline{L}^3_L\Phi_3\ell_{R}^3-Y_4^\ell\overline{L}^3_L\Phi_I\ell_{R}^I-Y_5^\ell\overline{L}^I_L\Phi_I\ell_{R}^3+h.c.,\\
\mathcal{L}_{Y_{\nu}}=&-Y_1^\nu\overline{L}^I_L\tilde{\Phi}_3\nu_{R}^I - Y_2^\nu \left(\overline{L}^I_L\sigma^1_{IJ}\tilde{\Phi}_1\nu_{R}^J + \overline{L}^I_L\sigma^3_{IJ}\tilde{\Phi}_2\nu_{R}^J\right)\nonumber\\
&-Y_3^\nu \overline{L}^3_L\tilde{\Phi}_3\nu_{R}^3 - Y_4^\nu\overline{L}^3_L\tilde{\Phi}_I\nu_{R}^I - Y_5^\nu\overline{L}^I_L\tilde{\Phi}_I\nu_{R}^3 + h.c.,
\end{align}
\label{equ:LYukawa}
\end{subequations}

\noindent where $I,J=1,2$, $\sigma^{1,2,3}$ are the Pauli matrices, $\tilde{\Phi}_i=i\sigma^2 \Phi_i^*$, and in general the matrices $Y_{1,\cdots,5}^{u,d,\ell,\nu}$ contain complex parameters.
The Lagrangian for the Majorana neutrino masses is given by
\begin{align}
    \mathcal{L}_M = -\frac{1}{2}M_1\nu_{R}^{I,T}C\nu_{R}^I - \frac{1}{2}M_3\nu_{R}^{3,T}C\nu_{R}^3,  \label{Majarona mass}
\end{align}
where $C$ is the charge conjugation matrix.
Then the mass matrices  for the fermions (Dirac mass matrix for the neutrino) are given by:
\begin{align}
M^{d,\ell}=\begin{pmatrix}
\epsilon^{d,\ell}&x^{d,\ell}m_1^{d,\ell} &x^{d,\ell}m_2^{d,\ell}\\[6pt]
x^{d,\ell}m_1^{d,\ell} & -2m_1^{d,\ell}+\epsilon^{d,\ell}&m_2^{d,\ell} \\[6pt]
x^{d,\ell}m_{4}^{d,\ell}&m_{4}^{d,\ell} &m_3^{d,\ell} \\[6pt]
\end{pmatrix},
M^{u,\nu}=\begin{pmatrix}
\epsilon^{u,\nu}&x^{u,\nu}m_1^{u,\nu} &x^{u,\nu}m_2^{u,\nu}\\[6pt]
x^{u,\nu}m_1^{u,\nu} & -2m_1^{u,\nu}+\epsilon^{u,\nu}&m_2^{u,\nu} \\[6pt]
x^{u,\nu}m_{4}^{u,\nu}&m_{4}^{u,\nu} &m_3^{u,\nu} \\[6pt]
\end{pmatrix}.
\label{equ:quark_mass}
\end{align}
where we have defined:
\begin{align}
&x^d=x^\ell=(x^u)^*=(x^\nu)^*=\frac{v_1e^{i\theta_1}}{v_2e^{i\theta_2}},\quad \epsilon^{d,\ell,u,\nu} = m_0^{d,\ell,u,\nu} + m_1^{d,\ell,u,\nu}, \nonumber \\
&\frac{m_1^{d,\ell}}{Y_2^{d,\ell}} = \frac{m_2^{d,\ell}}{Y_5^{d,\ell}} =\frac{m_4^{d,\ell}}{Y_4^{d,\ell}} = \left(\frac{m_1^{u,\nu}}{Y_2^{u,\nu}}\right)^* = \left(\frac{m_2^{u,\nu}}{Y_5^{u,\nu}}\right)^*=\left(\frac{m_4^{u,\nu}}{Y_4^{u,\nu}}\right)^* = \frac{v_2e^{i\theta_2}}{\sqrt{2}}, \nonumber \\
&\frac{m_3^{d,\ell}}{Y_3^{d,\ell}} = \frac{m_0^{d,\ell}}{Y_1^{d,\ell}} = \left(\frac{m_3^{u,\nu}}{Y_3^{u,\nu}}\right)^* = \left(\frac{m_0^{u,\nu}}{Y_1^{u,\nu}}\right)^* = \frac{v_3}{\sqrt{2}}.
\label{fermion mass matrices}
\end{align}
The above mass matrices can be diagonalized by bi-unitary transformations as
\begin{align}
U_{d(u,\ell)L}^{\dagger}M_{d(u,\ell)}U_{d(u,\ell)R}={\rm diag}(m_{d(u,e)},m_{s(c,\mu)},m_{b(t,\tau)}).\label{equ:fermion_transformation}
\end{align}
The quark mixing matrix $V_{\rm CKM}$ is then given by
\begin{align}
V_{\rm CKM}=U_{uL}^{\dagger}U_{dL}.\label{equ:ckm}
\end{align}
The Majorana mass matrix for the light neutrinos is obtained through the seesaw formula~\cite{Minkowski:1977sc,Yanagida:1979as,Mohapatra:1979ia,Glashow:1979nm}
\begin{align}
    M_{\nu}^{\rm Majorana} = M_{\nu}\tilde{M}^{-1}(M_\nu)^T
\end{align}
where $\tilde{M}={\rm diag}(M_1,M_1,M_3)$. Then the neutrino masses can be obtained by diagonalizing the light neutrino Majorana mass matrix,
\begin{align}
    U_\nu^T M_{\nu}^{\rm Majorana} U_\nu = {\rm diag}(m_{\nu_1}, m_{\nu_2}, m_{\nu_3}).
\end{align}
The mixing matrix in the lepton sector is then given by
\begin{align}
    U_{\rm PMNS} = U_{\ell\,L}^\dagger U_\nu .
\end{align}

\subsubsection*{Yukawa Couplings in the $S_3$ and the Mass Basis}
\label{section:Yukawa coupling matrices}
By expanding the Yukawa interactions in~\autoref{equ:LYukawa}, we are able to obtain the these couplings in the $S_3$ basis for the scalars given in~\autoref{equ:scalar_vevs} and the fermions given in ~\autoref{equ:fermion_S3}:
\begin{align}
    \mathcal{L}_{\rm Yukawa} =& -\phi_k\overline{u_L^i} \frac{\bar{Y}_{u,ij}^k}{\sqrt{2}} u_R^j - \phi_k\overline{d_L^i} \frac{\bar{Y}_{d,ij}^k}{\sqrt{2}}d_R^j - i \xi_k\overline{u_L^i}\frac{\tilde{Y}_{u,ij}^k}{\sqrt{2}}u_R^j - i \xi_k\overline{d_L^i}\frac{\tilde{Y}_{d,ij}^k}{\sqrt{2}}d_R^j\nonumber \\
    & - \omega_k^-\overline{d_L^i}\hat{Y}_{u,ij}^ku_R^j - \omega_k^+\overline{u_L^i}\hat{Y}_{d,ij}^k d_R^j + h.c.
\end{align}
where the Yukawa coupling matrices are given by
\begin{align}
K_{u,d}^1 &= \begin{pmatrix}
    0 & Y_2^{u,d} & Y_5^{u,d}\\
    Y_2^{u,d} & 0 & 0 \\
    Y_4^{u,d} & 0 & 0
\end{pmatrix}, &  K_{u,d}^2 &= \begin{pmatrix}
    Y_2^{u,d} & 0 & 0 \\
    0 & -Y_2^{u,d} & Y_5^{u,d} \\
    0 & Y_4^{u,d} & 0
\end{pmatrix}, & K_{u,d}^3 &= \begin{pmatrix}
    Y_1^{u,d} & 0 & 0 \\
    0 & Y_1^{u,d} & 0 \\
    0 & 0 & Y_3^{u,d}
\end{pmatrix}, \label{Yukawa_matrices }
\end{align}
with
\begin{align}
\bar{Y}_{u,d}^k &= K_{u,d}^k\,, & \tilde{Y}_{u,d}^k &= \mp K_{u,d}^k\,, & \hat{Y}_{u,d}^k &= \mp K_{u,d}^k\,.
\end{align}
The numerical values of~\autoref{Yukawa_matrices }  obatined from the best fit are given in~\autoref{C1}-\autoref{C6} in Appendix C.

In order to obtain the corresponding Yukawa coupling matrices in the mass basis (of both scalars and fermions), which are defined as
\begin{align}
\label{equ:Yukawa_mass_basis}
\mathcal{L}_{\rm Yukawa} = - h_k \overline{u_L^i}\frac{\mathbb{Y}_{u,ij}^k}{\sqrt{2}}u_R^j - h_k \overline{d_L^i}\frac{\mathbb{Y}_{d,ij}^k}{\sqrt{2}}d_R^j - h_k^- \overline{d_L^i}\hat{\mathbb{Y}}_{u,ij}^k u_R^j - h_k^+ \overline{u_L^i}\hat{\mathbb{Y}}_{d,ij}^k d_R^j + h.c.
\end{align}
we need to impose all the rotations in~\autoref{equ:scalar_higgs_basis}, \autoref{equ:scalar_neutral_mass_rotation}, \autoref{equ:scalar_charge_mass_rotation}, \autoref{equ:fermion_transformation}.
Assuming $\mathcal{Y}^k_{u,d} = (\bar{Y}^{1}_{u,d},\bar{Y}^{2}_{u,d},\bar{Y}^{3}_{u,d},i\tilde{Y}_{u,d}^{1},i\tilde{Y}_{u,d}^{2},i\tilde{Y}_{u,d}^{3})$ with $k=1,2,\cdots,6$, we have
\begin{align}
    \mathbb{Y}_{u(d),ij}^k &= (\mathbb{R}_H^\dagger R_0)_{\rho k}(U_{u(d)L}^\dagger \mathcal{Y}^\rho_{u(d)} U_{u(d)R})_{ij}, \\
    \hat{\mathbb{Y}}_{u,ij}^k &= (R_H^\dagger R_\pm)^*_{\rho k}(U_{dL}^\dagger \hat{Y}^\rho_{u}U_{uR})_{ij}, \\
    \hat{\mathbb{Y}}_{d,ij}^k &= (R_H^\dagger R_\pm)_{\rho k}(U_{uL}^\dagger \hat{Y}^\rho_dU_{dR})_{ij},
\end{align}
where
\begin{align}
\label{equ:rotation_combined}
    \mathbb{R}_H \equiv \begin{pmatrix}
        {\rm Re}(R_H) & -{\rm Im}(R_H) \\
        {\rm Im}(R_H) & {\rm Re}(R_H)
    \end{pmatrix}.
\end{align}
The Yukawa couplings in~\autoref{equ:Yukawa_mass_basis} will serve as the `standard' form for the calculation in the neutral meson mixing in the following section.

\subsection{Theoretical Constraints}
\label{sec:BFB_Unitarity}

The model itself will be constrained by several theoretical considerations. The Higgs potential should remain bounded from below (BFB).  Some necessary conditions for BFB have been worked out in Ref.~\cite{Emmanuel-Costa:2016vej}. However, it was pointed out in Ref.~\cite{Faro:2019vcd} that the parametrization of the three doublets is not accurate. In our work, we derived multiple necessary conditions by parametrizing the three Higgs doublets following Ref.~\cite{Faro:2019vcd},
 \begin{align}
\Phi_1 = r_1\left(\begin{array}{c}
    0\\
    1\end{array}\right),\quad
\Phi_2 = r_2\left(\begin{array}{c}
    \sin(\alpha_2)\\
    \cos(\alpha_2)e^{i\beta_2}
\end{array}\right),\quad
\Phi_3 = r_3e^{i\gamma}\left(\begin{array}{c}
    \sin(\alpha_3)\\
    \cos(\alpha_3)e^{i\beta_3}
\end{array}\right),
\end{align}
with $r_1= r\cos\theta$, $r_2=r\sin(\theta)\cos(\phi)$, $r_3=r\sin(\theta)\sin(\phi)$. Then we require the quartic term $V_4$ to be positive along some specific directions:
\begin{align}
   &V_4(\alpha_2=\frac{\pi}{2},\alpha_3=\beta_2=\beta_3=\gamma=\beta_4=\beta_7=0,r_1=r_2,r_3=0)>0, \nonumber \\
   &V_4(\alpha_2=\alpha_3=\beta_2=\beta_3=\gamma=\beta_4=\beta_7=0,r_1=r_2=0)>0, \nonumber\\
   &V_4(\beta_2=\frac{\pi}{2},\alpha_2=\alpha_3=\beta_3=\gamma=\beta_4=\beta_7=0,r_1=r_2,r_3=0)>0,  \nonumber\\
   &V_4(\alpha_2=\alpha_3=\beta_2=\beta_3=\gamma=\beta_4=\beta_7=0,r_2=r_3=0)>0, \nonumber\\
   &V_4(\alpha_3=\frac{\pi}{2},\alpha_2=\beta_2=\beta_3=\gamma=\frac{\pi}{4},\beta_4=\beta_7=0,r_1=0,r_2=r_3)>0, \nonumber\\
   &V_4(\alpha_2=\alpha_3=\beta_2=\beta_3=\frac{\pi}{2},\gamma=\beta_4=\beta_7=0,r_1=r_2=r_3)>0.
\end{align}
The corresponding necessary conditions from the above positivity conditions are given by
\begin{align}
   &\lambda_1 > 0, \nonumber\\
   &\lambda_8>0,\nonumber\\
   &\lambda_1-\lambda_2>0, \nonumber\\
   &\lambda_1+\lambda_3>0, \nonumber\\
   &\lambda_1+\lambda_3-\lambda_4+\lambda_5+\frac{1}{2}\lambda_6+\lambda_8 > 0, \nonumber\\
   &\lambda_1+\frac{1}{2}\lambda_5+\frac{1}{4}\lambda_6+\frac{1}{2}\lambda_7+\frac{1}{4}\lambda_8>0.
\end{align}
Since the above conditions are not sufficient, for specific parameter points we obtain during the fitting, we will also numerically check the minimum value of the quartic part of the potential explicitly to assure the BFB condition.

The perturbative unitarity constraints require
\begin{align}
\left|a_{i}^{\pm}\right|,\left|b_{i}\right| \leq 16 \pi, \quad i=1,2, \ldots, 6,
\end{align}
where $a_{i}^{\pm}$ and $b_{i}$ are the eigenvalues of the S-matrix of 3HDM with $S_3$ symmetry. In terms of the parameters in the potential, they can be expressed as~\cite{Das:2014fea}:
\begin{align}
    a_{1}^{\pm}=&\left(\lambda_{1}-\lambda_{2}+\frac{\lambda_{5}+\lambda_{6}}{2}\right) \nonumber \\
    &\qquad \pm \sqrt{\left(\lambda_{1}-\lambda_{2}+\frac{\lambda_{5}+\lambda_{6}}{2}\right)^{2}-4\left(\left(\lambda_{1}-\lambda_{2}\right)\left(\frac{\lambda_{5}+\lambda_{6}}{2}\right)-\lambda_{4}^{2}\right)},\nonumber \\
    a_{2}^{\pm}=&\left(\lambda_{1}+\lambda_{2}+2 \lambda_{3}+\lambda_{8}\right) \nonumber\\
    &\qquad \pm \sqrt{\left(\lambda_{1}+\lambda_{2}+2 \lambda_{3}+\lambda_{8}\right)^{2}-4\left(\lambda_{8}\left(\lambda_{1}+\lambda_{2}+2 \lambda_{3}\right)-2 \lambda_{7}^{2}\right)}, \nonumber \\
    a_{3}^{\pm}=&\left(\lambda_{1}-\lambda_{2}+2 \lambda_{3}+\lambda_{8}\right)\nonumber\\
    &\qquad \pm \sqrt{\left(\lambda_{1}-\lambda_{2}+2 \lambda_{3}+\lambda_{8}\right)^{2}-4\left(\lambda_{8}\left(\lambda_{1}-\lambda_{2}+2 \lambda_{3}\right)-\frac{\lambda_{6}^{2}}{2}\right)}, \nonumber \\
    a_{4}^{\pm}=&\left(\lambda_{1}+\lambda_{2}+\frac{\lambda_{5}}{2}+\lambda_{7}\right)\nonumber\\
    &\qquad \pm \sqrt{\left(\lambda_{1}+\lambda_{2}+\frac{\lambda_{5}}{2}+\lambda_{7}\right)^{2}-4\left(\left(\lambda_{1}+\lambda_{2}\right)\left(\frac{\lambda_{5}}{2}+\lambda_{7}\right)-\lambda_{4}^{2}\right)},\nonumber \\
    a_{5}^{\pm}=&\left(5 \lambda_{1}-\lambda_{2}+2 \lambda_{3}+3 \lambda_{8}\right) \nonumber \\
    & \pm \sqrt{\left(5 \lambda_{1}-\lambda_{2}+2 \lambda_{3}+3 \lambda_{8}\right)^{2}-4\left(3 \lambda_{8}\left(5 \lambda_{1}-\lambda_{2}+2 \lambda_{3}\right)-\frac{1}{2}\left(2 \lambda_{5}+\lambda_{6}\right)^{2}\right)},\nonumber  \\
    a_{6}^{\pm}=&\left(\lambda_{1}+\lambda_{2}+4 \lambda_{3}+\frac{\lambda_{5}}{2}+\lambda_{6}+3 \lambda_{7}\right) \nonumber \\
    &\pm \sqrt{(\lambda_1+\lambda_2+4\lambda_3+\frac{\lambda_5}{2}+\lambda_6+3\lambda_7)^2-4\left((\lambda_1+\lambda_2+4\lambda_3)(\frac{\lambda_5}{2}+\lambda_6+3\lambda_7)-9\lambda_4^2\right)} \nonumber \\
    b_{1}=& \lambda_{5}+2 \lambda_{6}-6 \lambda_{7},\nonumber  \\
    b_{2}=& \lambda_{5}-2 \lambda_{7}, \nonumber \\
    b_{3}=& 2\left(\lambda_{1}-5 \lambda_{2}-2 \lambda_{3}\right),\nonumber \\
    b_{4}=& 2(\lambda_1-\lambda_2-2\lambda_3),\nonumber \\
    b_{5}=& 2(\lambda_1+\lambda_2-2\lambda_3),\nonumber \\
    b_{6}=& \lambda_5-\lambda_6 .\label{unitarity10}
\end{align}

\section{Constraints from Neutral Meson Mixing and Neutron EDM}
\label{sec:fcnc-edm}
With the general structure in the Yukawa sector discussed in~\autoref{sec:yukawa_lagrangian}, the model will typically induce flavor changing neutral current (FCNC) processes. These will set stringent constraints on the model parameters, especially on the masses of the scalars. Here, we discuss the constraints from the neutral meson mixing induced by scalar exchange. In addition, this general structure of the Yukawa sector with CP violating phases will also provide new contributions to the electric dipole moment (EDM). In the second part of this section, we will discuss the constraints arising from the neutron EDM.

\subsection{Neutral Meson Mixing}
\label{section:FCNC}
\begin{figure}[htp]
    \centering
    \includegraphics[width=\textwidth]{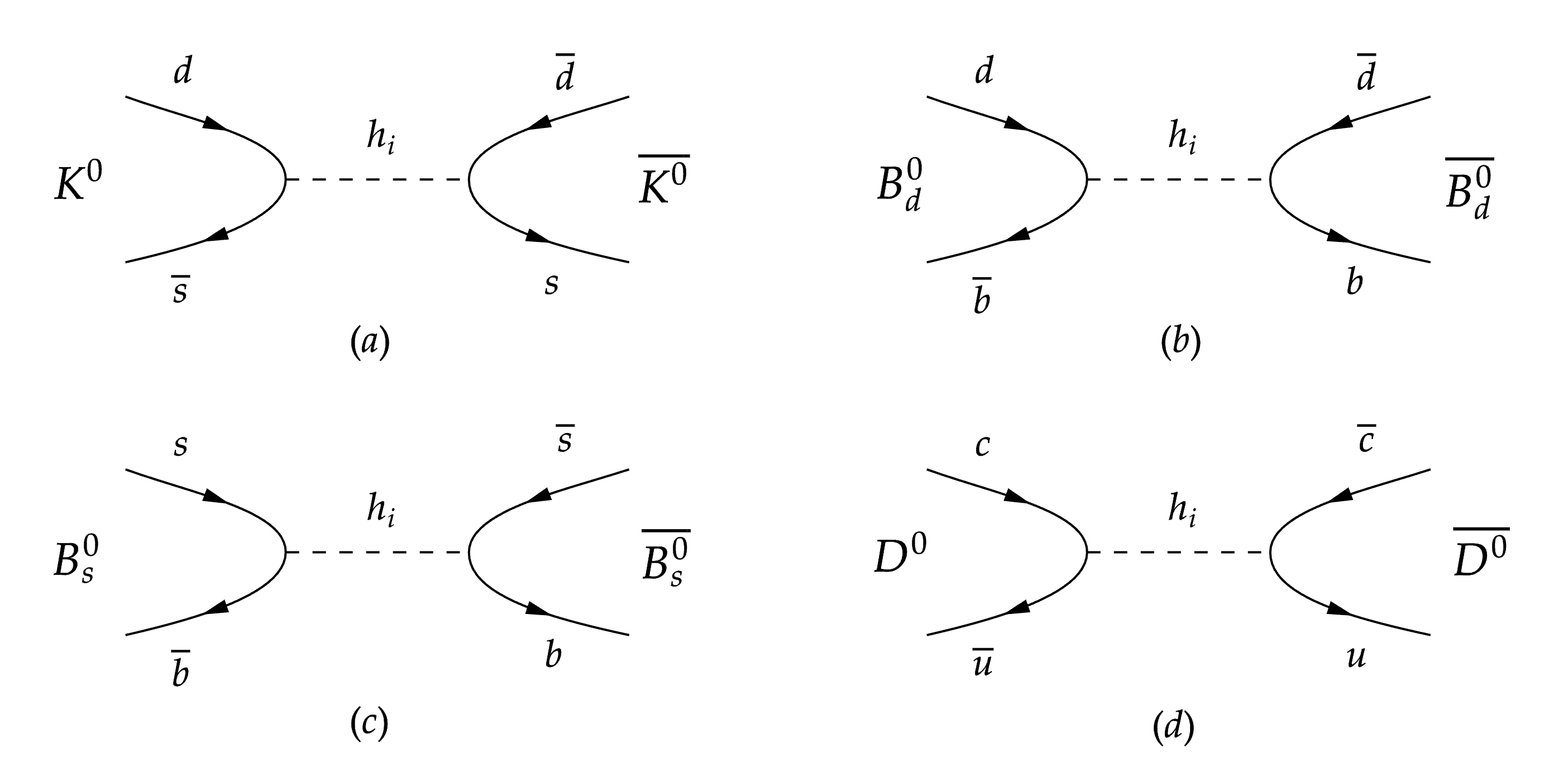}
    \caption{Feynman diagrams for various FCNC processes mediated by tree-level neutral Higgs boson exchange.}
    \label{fig:neutral meson mixing}
\end{figure}

The Feynman diagrams for the Higgs-mediated contributions to the neutral meson mixing (including $K^0$-$\overline{K^0}$, $B^0_d$-$\overline{B^0_d}$, $B^0_s$-$\overline{B^0_s}$ and $D^0$-$\overline{D^0}$) are shown in~\autoref{fig:neutral meson mixing}, where $h_i\,(i=1,\cdots,5)$ are the neutral scalars in the model.

The effective $\Delta F = 2$ Hamiltonian arising from these contributions is given by~\cite{Deshpande:1994en,Babu:2018uik,Babu:2009nn}:
\begin{align}
\mathcal{H}_{\mathrm{eff}}=-\frac{1}{2 m_{k}^{2}}\left(\bar{q}_{i}\left[\mathbb{Y}_{q,ij}^{k} \frac{1+\gamma_{5}}{2}+(\mathbb{Y}_{q}^{k\dagger})_{ij} \frac{1-\gamma_{5}}{2}\right] q_{j}\right)^{2}.
\end{align}
Here $q_{i, j}$ denotes the quark fields that the meson is composed of, and $\mathbb{Y}^k_q$'s are the Yukawa couplings of $q_{i}$, $q_{j}$ with scalar mass eigenstate $h_{k}$ with mass $m_k$ in~\autoref{equ:Yukawa_mass_basis}.
The corresponding transition matrix element can be written as
\begin{align}
\mathcal{M}_{12}^{\phi}=\left\langle\phi\left|H_{\mathrm{eff}}\right|\bar{\phi}\right\rangle=-\frac{f_{\phi}^{2} m_{\phi}}{2 m_{k}^{2}}&\left[-\frac{5}{24} \frac{m_{\phi}^{2}}{\left(m_{q_{i}}+m_{q_{j}}\right)^{2}}\left(\mathbb{Y}_{q,ij}^{k^{2}}+\mathbb{Y}_{q, ji}^{k^{* 2}}\right) \cdot B_{2} \cdot \eta_{2}(\mu)\right. \nonumber\\
&\left.+\mathbb{Y}_{q,ij}^{k} \mathbb{Y}_{q,ji}^{k^{*}}\left(\frac{1}{12}+\frac{1}{2} \frac{m_{\phi}^{2}}{\left(m_{q_{i}}+m_{q_{j}}\right)^{2}}\right) \cdot B_{4} \cdot \eta_{4}(\mu)\right],
\label{FCNC equation}
\end{align}
where $\phi$ denotes the neutral mesons $\left(K^{0}, B_{d}^{0}, B_{s}^{0}, D^{0}\right)$. We use the modified vacuum saturation and factorization approximation to parameterize the matrix elements for our numerical study:
\begin{align}
\left\langle\phi\left|\bar{f}_{i}\left(1 \pm \gamma_{5}\right) f_{j} \bar{f}_{i}\left(1 \mp \gamma_{5}\right) f_{j}\right| \bar{\phi}\right\rangle&=f_{\phi}^{2} m_{\phi}\left(\frac{1}{6}+\frac{m_{\phi}^{2}}{\left(m_{q_{i}}+m_{q_{j}}\right)^{2}}\right) \cdot B_{4}, \\
\left\langle\phi\left|\bar{f}_{i}\left(1 \pm \gamma_{5}\right) f_{j} \bar{f}_{i}\left(1 \pm \gamma_{5}\right) f_{j}\right| \bar{\phi}\right\rangle&=-\frac{5}{6} f_{\phi}^{2} m_{\phi} \frac{m_{\phi}^{2}}{\left(m_{q_{i}}+m_{q_{j}}\right)^{2}} \cdot B_{2} .
\end{align}
The values of $B_2$, $B_4$ as well as $m_\phi$ and $f_\phi$ for different systems are summarized in~\autoref{tab:coeff_mixing}.
$\eta_2(\mu)$ and $\eta_4(\mu)$ are QCD correction factors of the Wilson coefficients of the effective $\Delta F=2$ Hamiltonian in going from the heavy scalar mass scale $m_{s}$ to the hadronic scale $\mu$.
These factors can be computed as follows. The $\Delta F=2$ effective Hamiltonian has the general form
\begin{align}
\mathcal{H}_{\mathrm{eff}}^{\Delta F=2}=\sum_{i=1}^{5} C_{i} Q_{i}+\sum_{i=1}^{3} \tilde{C}_{i} \tilde{Q},
\end{align}
where
\begin{align}
Q_{1}&=\bar{q}_{iL}^{\alpha} \gamma_{\mu} q_{jL}^{\alpha} \bar{q}_{iL}^{\beta} \gamma^{\nu} q_{jL}^{\beta},& Q_{2}&=\bar{q}_{iR}^{\alpha} q_{jL}^{\alpha} \bar{q}_{i R}^{\beta} q_{j R}^{\beta},& Q_{3}=\bar{q}_{iR}^{\alpha} q_{j L}^{\beta} \bar{q}_{i R}^{\beta} q_{jL}^{\alpha},\nonumber \\
Q_{4}&=\bar{q}_{iR}^{\alpha} q_{jL}^{\alpha} \bar{q}_{i L}^{\beta} q_{j R}^{\beta}, & Q_{5}&=\bar{q}_{iR}^{\alpha} q_{j L}^{\beta} \bar{q}_{i L}^{\beta} q_{j R}^{\alpha},
\end{align}
$\tilde{Q}_{1,2,3}$ can be obtained from $Q_{1,2,3}$ by interchanging $L \leftrightarrow R$. The new physics scale $m_{s}$ is taken to be $1\ \mathrm{TeV}$. The evolution of the Wilson coefficients from $m_{s}$ down to the hadron scale $\mu$ can be obtained from
\begin{align}
C_{r}(\mu)=\sum_{i} \sum_{s}\left(b_{i}^{(r, s)}+\eta c_{i}^{(r, s)}\right) (\eta)^{a_{i}} C_{s}\left(M_{s}\right).
\label{equ:RGRunning_WC}
\end{align}
Here $\eta=\alpha_{s}\left(m_{s}\right) / \alpha_{s}\left(m_{t}\right)$. The magic numbers $a_{i}, b_{i}^{(r, s)}$ and $c_{i}^{(r, s)}$ are taken from Ref.~\cite{Ciuchini:1998ix} for the $K$ meson system, from Ref.~\cite{Becirevic:2001jj} for the $B_{d, s}$ meson system and from Ref.~\cite{UTfit:2007eik} for the $D$ meson system. With $m_{s}=1\ \mathrm{TeV}, m_{t}\left(m_{t}\right)=163.6\ \mathrm{GeV}$ and $\alpha_{s}\left(m_{Z}\right)=0.118$, we have $\eta=\alpha_{s}(1\ \mathrm{TeV}) / \alpha_{s}\left(m_{t}\right)=0.8167$. At the mass scale of the heavy scalars, only operators $Q_2$ and $Q_4$ are generated. We obtain the coefficients of different operators at low scales according  to~\autoref{equ:RGRunning_WC}, and hence the $\eta_2(\mu)$ and $\eta_4(\mu)$ for different systems, which are also shown in~\autoref{tab:coeff_mixing}.

\begin{table}[!tb]
    \centering
    \begin{tabular}{|c|c|c|c|c|c|c|}
    \hline\hline
    System & $B_2$ & $B_4$ & $m_\phi$ & $f_\phi$ & $\eta_2(\mu)$ & $\eta_4(\mu)$ \\
    \hline
    $K^0$ & 0.66 & 1.03 & 498 MeV & 160 MeV & 2.54 & 4.81\\
    \hline
    $B_d^0$ & 0.82 & 1.16 & 5.28 GeV  & 240 MeV & 2.00 & 3.12\\
    \hline
    $B_s^0$ & 0.82 & 1.16 & 5.37 GeV & 295 MeV & 2.00 & 3.12\\
    \hline
    $D^0$ & 0.82 & 1.08 & 1.86 GeV & 200 MeV & 2.31 & 3.99\\
    \hline\hline
    \end{tabular}
    \caption{The value of $B_2$, $B_4$, $m_\phi$, $f_\phi$ and $\eta_2(\mu)$, $\eta_4(\mu)$ in different systems~\cite{Babu:2018uik,Babu:2009nn,Bauer:2009cf}.}
    \label{tab:coeff_mixing}
\end{table}

The constraints from the neutral meson mixing are usually quoted as upper bounds on the mass difference and/or the CP violation parameter which are defined as:
\begin{align}
\Delta m_\phi &= 2\Re \mathcal{M}_{12}^\phi, \\
\epsilon_\phi &= \frac{\Im \mathcal{M}_{12}^\phi}{\sqrt{2}\Delta m_\phi}.
\end{align}
For different meson systems, we have the following constraints~\cite{Babu:2018uik}:
\begin{align}
    \text{$K^0$-$\overline{K^0}$}:&\quad \Delta m_K\lesssim 3.484\times10^{-15}\,\rm GeV,\quad |\epsilon_K|\lesssim2.23\times10^{-3},\\
    \text{$B_d^0$-$\overline{B_d^0}$}:&\quad \Delta m_{B_d}\lesssim 3.12\times10^{-13}\,\rm GeV,\\
    \text{$B_s^0$-$\overline{B_s^0}$}:&\quad \Delta m_{B_s}\lesssim 1.17\times10^{-11}\,\rm GeV,\\
    \text{$D^0$-$\overline{D^0}$}:&\quad \Delta m_D\lesssim 6.25\times10^{-15}\,\rm GeV~.
\end{align}

\subsection{Neutron Electric Dipole Moment}
\begin{figure}
    \centering
    \includegraphics[width=\textwidth]{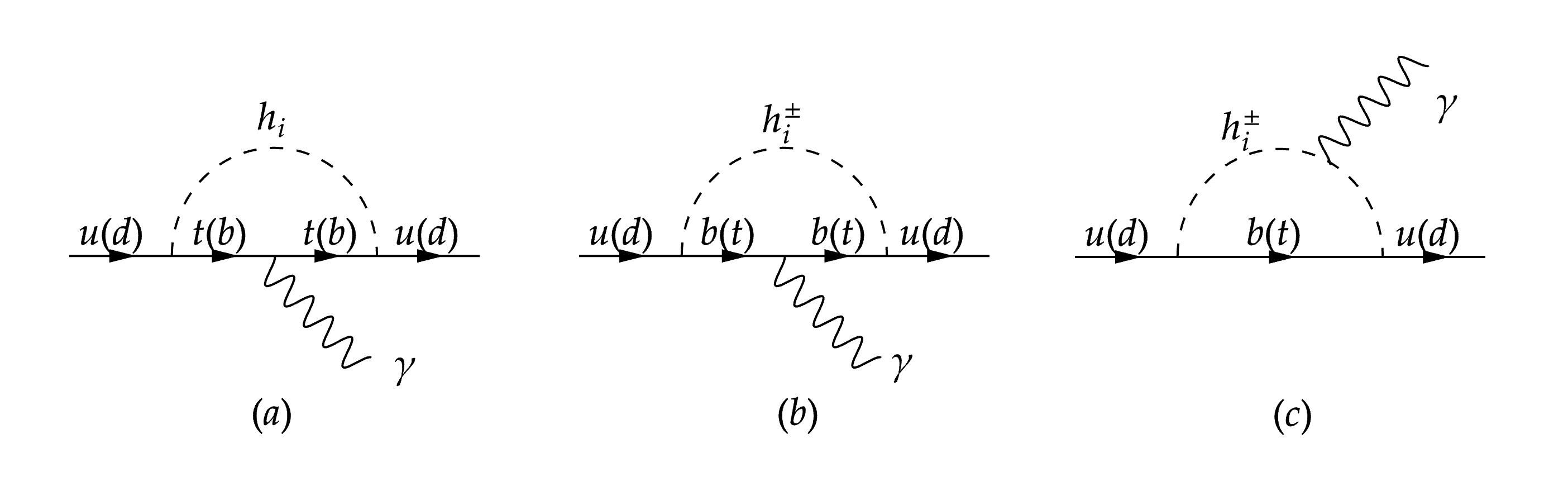}
    \caption{New contributions to the quark EDM through neutral scalar (a) and through charged scalar (b), (c).}
    \label{fig:EDM_Feynman}
\end{figure}

The current measurement of the neutron EDM puts strong constraints on the model parameters. The latest measurement requires that~\cite{Abel:2020pzs}
\begin{align}
    |d_n|<1.8\times 10^{-26}\, {\rm e}\cdot{\rm cm}~.
\end{align}
In our $S_3$ model, there will be new contributions to the electric dipole moments (EDM) of the fermions from either the neutral scalars or the charged scalar exchange. The main contributions are from the one-loop diagrams presented in~\autoref{fig:EDM_Feynman}.
In the most general case, for a given fermion $f$, we assume that it has the following Yukawa couplings with fermions $\psi_i$ and heavy scalar $\phi_j$ with mass $m_i$ and $m_j$ respectively:
\begin{align}
\mathcal{L}_{\psi f \phi}=L^f_{i j}\left(\bar{\psi}_{i} P_{L} f\right) \phi_{j}+R^f_{i j}\left(\bar{\psi}_{i} P_{R} f\right) \phi_{j}+h.c.,
\end{align}
where $P_{R, L}=\frac{1}{2}\left(1 \pm \gamma^{5}\right)$ and $L^f_{i j}$ and $R^f_{i j}$ are the Yukawa couplings related to fermion $f$ which can be obtained in our model by matching to that in~\autoref {equ:Yukawa_mass_basis}. With these general Yukawa interactions, the one-loop contributions to the EDM of fermion $f$ are given by~\cite{Ibrahim:2007fb}
\begin{align}
    {d_{f}}/e=\left(\frac{m_{i}}{16 \pi^{2} m_{j}^{2}}\right) \operatorname{Im}\left[\left(R_{i j}\right)^{*} L_{i j}\right]\left[Q_{i}A\left(\frac{m_i^2}{m_j^2}\right)+Q_{j}B\left(\frac{m_i^2}{m_j^2}\right)\right],
    \label{equu:EDM_fermion}
    \end{align}
where $Q_i$ and $Q_j$ are the electric charge of $\psi_i$ and $\phi_j$ respectively, and the loop functions $A(r)$ and $B(r)$ are defined as
\begin{align}
    A(r) &= \frac{1}{2(1-r)^2}\left(3-r + \frac{2\log r}{1-r}\right), \\
    B(r) &= \frac{1}{2(1-r)^2}\left(1+r + \frac{2r\log r}{1-r}\right).
\end{align}
For the neutron EDM, we also need to consider the contribution from chromoelectric dipole moments (CEDM) $d_f^c$. The CEDM can be obtained as~\cite{Iltan:2001vg}
\begin{align}
    d_f^c/g_s = \left.d_f/e\right|_{Q_i\to1,Q_j\to0}~.
\end{align}
Then the total contributions to the neutron EDM can be expressed as~\cite{Bertolini:2019out}
\begin{align}
d_{n} \simeq 0.32 d_{d}-0.08 d_{u}+e\left(0.12 d_{d}^c-0.12 d_{u}^c-0.006 d_{s}^c\right).
\end{align}

\section{Fitting Results}
\label{sec:Results}
\subsection{Analytical Approximation}

While we have carried out numerical fits to the fermion masses and mixing parameters arising from the mass matrix~\autoref{equ:quark_mass}, we have used approximate analytic results as our starting point.  We describe this approximation here. Note that if we ignore $x^d$ and $\epsilon^{d,\ell,u}$ in~\autoref{equ:quark_mass}, the first family quark and lepton masses would vanish, and their mixings with the second and third family would also vanish.  If we further assume that $m_3^f \gg m_{1,2,4}^f$ for $f=(d,\ell,u)$, one would realize the hierarchical structure of the three family masses.  In this approximation, only the third family fermion acquires a nonzero mass in the $S_3$ symmetric limit, with the lighter families acquiring their masses upon $S_3$ symmetry breaking.

The mass matrices given in~\autoref{equ:quark_mass} can be diagonalized sequentially by three unitary rotations, which are given, for a general matrix $M^f$, as
\begin{align}
U_{23}^{L,R} = \begin{pmatrix}
    1 & 0& 0\\[6pt]
    0 & 1 & \epsilon_{23}^{L,R}\\[6pt]
    0& -(\epsilon_{23}^{L,R})^*& 1\\[6pt]
\end{pmatrix},
U_{12}^{L,R}= \begin{pmatrix}
    1 & \epsilon_{12}^{L,R}& 0\\[6pt]
     -(\epsilon_{12}^{L,R})^* & 1 & 0\\[6pt]
    0&0& 1\\[6pt]
\end{pmatrix},
U_{13}^{L,R}= \begin{pmatrix}
    1 & 0& \epsilon_{13}^{L,R}\\[6pt]
     0 & 1 & 0\\[6pt]
    -(\epsilon_{13}^{L,R})^*&0& 1\\[6pt]
\end{pmatrix}.
\end{align}
with
\begin{align}
 \epsilon_{23}^L=\frac{-m_2}{m_3}, \quad \epsilon_{23}^R=\frac{-m_4}{m_3},
\end{align}
\begin{align}
    \epsilon_{12}^L=(\epsilon_{12}^R)^*=-x\frac{1-\frac{m_1m_3}{m_2m_4}}{1+2\frac{m_1m_3}{m_2m_4}},
\end{align}
\begin{align}
    \epsilon_{13}^L=-\frac{xm_2}{m_3}, \quad (\epsilon_{13}^R)^*=-\frac{xm_4}{m_3}.
\end{align}
With the approximation $|\frac{m_1m_3}{m_2m_4}|\ll 1$ applied to all sectors, we obtain:
\begin{align}
    |V_{us}|\simeq |x^d-(x^d)^*|, \quad V_{ub}\simeq \left|x^d\frac{m_2^d}{m_3^d}\right|,\quad
    |V_{cs}|\simeq 1, \quad |V_{cb}|\simeq \left|-\frac{m_2^u}{m_3^u}+\frac{m_2^d}{m_3^d}\right|,
\end{align}
\begin{align}
    m_b\simeq |m_3^d|, \quad m_s\simeq \left|-\frac{m_2^dm_4^d}{m_b}\right|,
    \quad m_d\simeq \left|\epsilon^d+(x^d)^2\,\frac{m_2^dm_4^d}{m_b}\right|,
\end{align}
\begin{align}
    m_t\simeq| m_3^u|, \quad m_c\simeq \left|-\frac{m_2^um_4^u}{m_u}\right|,
    \quad m_u\simeq \left|\epsilon^u+(x^d)^{*2}\,\frac{m_2^um_4^u}{m_b}\right|.
\end{align}

We set up the initial range of parameters by using the above analytical expressions, which point to the approximate range of the parameters, although the exact ranges are obtained numerically.

\label{sec:results}
\subsection{Fit to quark masses and CKM mixing angles}
\begin{table}[!htp]
    \centering
    \begin{tabular}{|c|c|c|c|}
    \hline
    \multirow{2}{5em}{Observables}&\multicolumn{3}{c|}{3HDM with $S_3$}\\
    \cline{2-4}
    &Input&Best Fit&Pull\\
    \hline
    $m_u(\mathrm{GeV})$&0.0011$\pm$0.000375&0.0008731&0.61 \\
    $m_c(\mathrm{GeV})$&0.532$\pm$0.002&0.5319&0.05\\
    $m_t(\mathrm{GeV})$&150.7$\pm$0.5&150.85&0.30 \\
    $m_d(\mathrm{GeV})$&0.0025$\pm$0.000325&0.0024771&0.07 \\
    $m_s(\mathrm{GeV})$&0.047$\pm$0.008&0.0443&0.34\\
    $m_b(\mathrm{GeV})$&2.43$\pm$0.025&2.4466&0.66 \\
    $|V_{ud}|$&0.9737$\pm$0.00014&0.974049&2.49\\
    $|V_{us}|$&0.2245$\pm$0.0008&0.22630&2.25 \\
    $|V_{ub}|$&0.00382$\pm$0.00024&0.003878&0.24\\
    $|V_{cd}|$&0.221$\pm$0.004&0.2262&1.30 \\
    $|V_{cs}|$&0.987$\pm$0.011&0.9732&1.25 \\
    $|V_{cb}|$&0.041$\pm$0.0014&0.03972& 0.91\\
    $|V_{td}|$&0.008$\pm$0.0003&0.00808&0.27 \\
    $|V_{ts}|$&0.0388$\pm$0.0011&0.03908&0.25 \\
    $|V_{tb}|$&1.013$\pm$0.03&0.9992&0.46 \\
    $J_{CP}^{CKM}$&$(3\pm0.12)\times10^{-5}$&$3.051\times10^{-5}$&0.42 \\
    \hline
    $\chi^2$&$-$&$-$&16.94 \\
    \hline
    \end{tabular}
    \caption{Inputs from the quark sector and corresponding best fit values of the observables along with their pulls at scale $\mu$ = 1 TeV. Observables \{$|V_{ud}|,|V_{us}|,|V_{ub}|,|V_{cd}|,|V_{cs}|,|V_{cb}|,|V_{td}|,|V_{ts}|,|V_{tb}|$\} are the absolute value of elements of the CKM matrix and $J_{CP}^{CKM}$ is the Jarlskog invariant related to the phase of the CKM matrix.}
    \label{tab:quark_measurements}
\end{table}
The fitting of the model parameters for the quark sector and the lepton sector are performed separately. In the first step, we obtain the complex parameters $\{x, m_1^{d,u}, m_2^{d,u}, m_3^{d,u},\newline  m_4^{d,u}, \epsilon^{d,u}\}$ in the quark sector by fitting the quark masses and CKM matrix elements by fitting~\autoref{equ:quark_mass} and~\autoref{equ:ckm} to the measured values of these parameters~\cite{Babu:2009fd,ParticleDataGroup:2020ssz} shown in~\autoref{tab:quark_measurements}. Since we consider the new physics which might contain new scalars at several TeV, the fitting is effectively performed at the TeV scale. The quark masses and the elements of CKM matrix are hence taken at 1 TeV scale as input~\cite{Babu:2009fd}. The results of the complex parameters in the quark sector of the best-fit point are shown in~\autoref{equ:quark_inputs} and the corresponding value of the observables (quark masses and CKM matrix) are also listed in~\autoref{tab:quark_measurements}\footnote{In the quark sector and the lepton sector discussed below, we use {\tt MultiNest}~\cite{Feroz:2008xx,Feroz:2013hea} and its {\tt Python} interface {\tt PyMultiNest}~\cite{Buchner:2014nha} to scan the parameter space and obtain the best-fit parameters with the likelihood constructed from the theoretical predictions, observations and corresponding errors.}. The best fit differs from the input value the largest in $|V_{ud}|$ and $|V_{us}|$ with the fit for $|V_{ud}|$ differing by 0.0003, and $|V_{us}|$ by 0.0018.  Since currently there appears to be an anomaly in the extraction of the Cabibbo angle (the Cabibbo anomaly), we feel that the total $\chi^2$ found in the fit is acceptable.
\begin{align}
x&=1.01\times10^{-1} + 1.59\times 10^{-1}i, \nonumber \\
\epsilon^d &= -6.15\times 10^{-5} + 4.34\times 10^{-4}i\,{\rm GeV}, &&\quad \epsilon^u = -6.08\times10^{-5}-9.18\times10^{-4}i\,{\rm GeV},\nonumber \\
m_1^d&=3.09\times10^{-2}+5.23\times10^{-3}i\,{\rm GeV}, &&\quad m_1^u = 4.46\times10^{-4}-9.49\times10^{-4}i\,{\rm GeV},\nonumber \\
m_2^d&=-2.10\times10^{-2} + 2.01\times10^{-2}i\,{\rm GeV}, &&\quad m_2^u = 1.77\times 10^{0} + 2.08\times10^{0}i\,{\rm GeV},\nonumber \\
m_3^d&=-7.08\times10^{-2} -2.12\times10^{0}i\,{\rm GeV}, &&\quad m_3^u = 1.39\times10^{2}-5.00\times10^{1}i\,{\rm GeV},\nonumber \\
m_4^d&=4.96\times10^{-1} - 1.09\times10^{0}i\,{\rm GeV}, &&\quad m_4^u = -9.62\times10^{-1} + 2.84\times10^{1}i\,{\rm GeV}.
\label{equ:quark_inputs}
\end{align}
With the values of the above parameters, we can obtain the corresponding values of the quark mass matrices $M_{d,u}$ shown in~\autoref{up quark mass matrix result}-\autoref{down quark mass matrix result} and transformation matrices $U_{(d,u)(L,R)}$ shown in~\autoref{quarktransformation1}-\autoref{quarktransformation4}.
\begin{align}
    \frac{M_u}{\rm GeV}=\begin{pmatrix}
    (-0.608-9.180i)\times10^{-4}&(-1.061-1.669i)\times10^{-4}& (5.103-0.720i)\times10^{-1}\\[6pt]
    (-1.061-1.669i)\times10^{-4}&(-9.521+9.810i)\times10^{-4}&(1.772+2.081i)\times10^0 \\[6pt]
    (4.422+3.021i)\times10^0&(-0.0962+2.839i)\times10^1&(1.393-0.500i)\times10^2
    \end{pmatrix},\label{up quark mass matrix result}
    \end{align}
    \begin{align}
    \frac{M_d}{\mathrm{GeV}}=\begin{pmatrix}
    -(0.615-4.342i)\times10^{-4} & (2.291+5.449i)\times10^{-3}&(-5.318-1.318i)\times10^{-3}\\[6pt]
    (2.291+5.449i)\times10^{-3}&(-6.188-1.001i)\times10^{-2}&(-2.101+2.007i)\times10^{-2} \\[6pt]
    (2.231-0.308i)\times10^{-1}&(0.496-1.086i)\times10^{0}&(-0.0708-2.122i)\times10^0
    \end{pmatrix},\label{down quark mass matrix result}
\end{align}
\begin{align}
    U_{uL}=\begin{pmatrix}
    0.983&0.0995-0.157i&0.00328+0.000681i\\[6pt]
    -0.0995-0.157i&0.982&0.00628+0.0166i\\[6pt]
    0.0000172+0.00000132i&-0.00639+0.0169i &0.9998 \label{quarktransformation1}
    \end{pmatrix},
    \end{align}
    \begin{align}
    U_{dL}=\begin{pmatrix}
    0.986&-0.125-0.107i&-0.000274-0.000985i\\[6pt]
    0.125-0.107i&0.986&-0.0101-0.0195i\\[6pt]
    -0.000554-0.00449i&0.00985-0.0192i&0.9997
    \end{pmatrix},
    \end{align}
    \begin{align}
    U_{uR}=\begin{pmatrix}
    0.0653+0.980i&0.0431+0.178i&0.0293-0.020i\\[6pt]
    -0.164-0.089i&0.913+0.309i&-0.00638-0.189i \\[6pt]
    -0.000101+0.000148i&0.124-0.146i&0.924+0.332i
    \end{pmatrix},
    \end{align}
    \begin{align}
    U_{dR}=\begin{pmatrix}
    -0.138-0.957i&0.113-0.210i&0.0911+0.0126i\\[6pt]
    0.0882-0.141i&-0.816+0.259i&0.203+0.444i \\[6pt]
    -0.179+0.0739i&0.453+0.0645i&-0.029+0.867i \label{quarktransformation4}
    \end{pmatrix}.
\end{align}

\subsection{Lepton Masses Fit}

\begin{table}[!htp]
    \centering
    \resizebox{15cm}{!}{
    \begin{tabular}{|c|c|c|c|c|c|c|}
    \hline
    \multirow{2}{5em}{Observables}&\multicolumn{3}{c|}{Normal Ordering}&\multicolumn{3}{c|}{Inverted Ordering}\\
    \cline{2-7}
    &Input&Best Fit&Pull&Input&Best Fit&Pull\\
    \hline
    $m_e(\mathrm{GeV})$&0.000511$\pm$0.00000511&0.0005109&0.02&0.000511$\pm$ 0.00000511&0.0005181&1.39 \\
    $m_\mu(\mathrm{GeV})$&0.105658$\pm$0.00105658&0.1052673&0.37&0.1056$\pm$ 0.00105 &0.106495&0.85\\
    $m_\tau(\mathrm{GeV})$&1.77$\pm$0.0177&1.763&0.40&1.777$\pm$ 0.0177&1.759&1.02\\
    $\Delta m_{21}^2/10^{-5}\mathrm{(eV^2)}$&7.42$\pm$0.21&7.842&2.01&7.42$\pm$ 0.21&7.177&1.16\\
    $\Delta m_{3\ell}^2/10^{-3}\mathrm{(eV^2)}$&2.517$\pm$0.027&2.508&0.33&-2.498$\pm$ 0.028&-2.472&0.93\\
    $|U_{e1}|$&0.825$\pm$0.0072&0.8243&0.10&0.825$\pm$ 0.0073&0.8243&0.10\\
    $|U_{e2}|$&0.545$\pm$0.0109&0.5462&0.11&0.545$\pm$0.011&0.5457&0.06 \\
    $|U_{e3}|$&0.149$\pm$0.0021&0.14896&0.02&0.150$\pm$0.0021&0.1507&0.33 \\
    $|U_{\mu1}|$&0.271$\pm$0.0436&0.2810&0.23&0.390$\pm$0.0473&0.4290&0.82 \\
    $|U_{\mu2}|$&0.604$\pm$0.0299&0.5937&0.34&0.535$\pm$0.0333&0.5056&0.88 \\
    $|U_{\mu3}|$&0.749$\pm$0.0118&0.7540& 0.42&0.750$\pm$0.113&0.7486&0.01\\
    $|U_{\tau1}|$&0.495$\pm$0.0373&0.4915&0.09&0.409$\pm$0.0413&0.3694&0.96 \\
    $|U_{\tau2}|$&0.581$\pm$0.0271&0.5910&0.37&0.646$\pm$0.0284&0.6683&0.79 \\
    $|U_{\tau3}|$&0.646$\pm$0.0137&0.6397&0.99&0.645$\pm$0.0131&0.6467&0.13 \\
    \hline
    $\chi^2$&$-$&$-$&5.93&-&-&9.04 \\
    \hline
    \end{tabular}
    }
    \caption{Leptons inputs and corresponding best fit values of the observable along with their pulls at scale $\mu$ = 1 TeV. Observables \{$|U_{e1}|,|U_{e2}|,|U_{e3},|U_{\mu1}|,|U_{\mu2}|,|U_{\mu3}|,|U_{\tau1}|,|U_{\tau2}|,|U_{\tau3}|$\} are the absolute values of elements of the PMNS matrix. Note that $\Delta m_{3\ell}^2 \equiv \Delta m_{31}^2 > 0$ for normal ordering of neutrino masses and $\Delta m_{3\ell}^2 \equiv \Delta m_{32}^2 < 0$ for inverted ordering. }
    \label{table:lepton masses fit table}
\end{table}

The second step is fitting the lepton sector by adopting the result of the ratio $x=v_1e^{i\theta_1}/v_2e^{i\theta_2}$ from the quark sector to the charged lepton masses~\cite{Babu:2009fd}, the squared mass differences of the neutrinos corresponding to normal ordering as well as inverted ordering, and the PMNS matrix~\cite{Esteban:2020cvm} listed in~\autoref{table:lepton masses fit table}. We see that the fits to both normal mass ordering and inverted mass ordering are very good, with the former case giving a smaller $\chi^2$. In this way, the free parameters in the lepton sector are $\{m_1^{\ell,\nu},m_2^{\ell,\nu},m_3^{\ell,\nu},m_4^{\ell,\nu},\epsilon^{\ell,\nu},M_1,M_3\}$. Similar to the quark sector, we use the charged lepton masses at 1 TeV scale. The results of the free parameters in the lepton sector of the best-fit point of normal ordering are found to be:
\begin{align}
\epsilon^\ell&=-1.76\times10^{-4}+9.16\times10^{-5}i\,{\rm GeV},&&\quad \epsilon^\nu = -1.47\times10^{-1}+2.03\times10^{-1}i\,{\rm GeV},\nonumber \\
m_1^\ell&=9.76\times10^{-4}-2.90\times10^{-3}i\,{\rm GeV},&&\quad m_1^\nu = 5.69\times10^{-2}-7.48\times10^{-2}i\,{\rm GeV},\nonumber\\
m_2^\ell&=-9.81\times10^{-2}-1.64\times10^{-1}i\,{\rm GeV},&&\quad m_2^\nu = -6.42\times10^{-6}+1.07\times10^{-5}i\,{\rm GeV},\nonumber \\
m_3^\ell&=1.27\times10^{0}-7.51\times10^{-1}i\,{\rm GeV},&&\quad m_3^\nu = -4.80\times10^1+3.02\times10^1i\,{\rm GeV},\nonumber\\
m_4^\ell&=-4.12\times10^{-1}+8.45\times10^{-1}i\,{\rm GeV},&&\quad m_4^\nu = -1.22\times10^{-1}+4.43\times10^{-1}i\,{\rm GeV},\nonumber\\
M_1&=-6.75\times10^9+1.52\times10^9i\,{\rm GeV},&&\quad M_3=9.24\times10^{12}+1.84\times10^{14}i\,{\rm GeV}.
\label{equ:lepton_inputs normal}
\end{align}
With the above best-fit parameters, we have the numerical value of lepton mass matrices $M^{\ell,\nu}$ in~\autoref{lepton mass matrix result}-\autoref{neutrino mass matrix result} and the mixing matrices $U_{\ell(L,R)}$ and $U_\nu$ in~\autoref{lepton transformation1}-\autoref{lepton transformation2} corresponding to the normal ordering:
\begin{align}
    \frac{M_e}{\mathrm{GeV}}=\begin{pmatrix}
    (-1.763+0.916i)\times10^{-4}&(5.606-1.377i)\times10^{-4}&(1.624-3.221i)\times10^{-2}\\[6pt]
    (5.606-1.377i)\times10^{-4}&(-2.129+5.896i)\times10^{-3}&(-0.981-1.643i)\times10^{-1} \\[6pt]
    (-1.761+0.198i)\times10^{-1}&(-4.118+8.449i)\times10^{-1}&(1.265-0.751i)\times10^{0} \\[6pt]
    \end{pmatrix},\label{lepton mass matrix result}
    \end{align}
    \begin{align}
    \frac{M_{\nu}^{\rm Majorana}}{10^{-12}\mathrm{GeV}}=\begin{pmatrix}
    0.895+8.989i&-1.547-0.846i&1.947+0.193i\\[6pt]
    -1.547-0.846i&2.116+27.654i&12.333+26.462i \\[6pt]
    1.947+0.193i&12.333+26.462i&7.121+11.614i \\[6pt]
    \end{pmatrix},\label{neutrino mass matrix result}
    \end{align}
    \begin{align}
    U_{eL}=\begin{pmatrix}
    0.983&0.0876+0.163i&0.0144-0.00940i\\[6pt]
    -0.0881+0.164i&0.978&0.00164-0.0915i\\[6pt]
    0.000999-0.00144i&-0.00134-0.0931i &0.996 \\[6pt]
    \end{pmatrix},\label{lepton transformation1}
    \end{align}
    \begin{align}
    U_{eR}=\begin{pmatrix}
    -0.880+0.438i&-0.0101+0.157i&-0.0995-0.0112i\\[6pt]
    0.152+0.108i&-0.762+0.321i&-0.233-0.477i\\[6pt]
    -0.000656+0.0086i&-0.299+0.450i &0.723+0.430i \\[6pt]
    \end{pmatrix},
    \end{align}
    \begin{align}
        U_\nu=\begin{pmatrix}
        0.767&0.642&-0.00550-0.00112i\\[6pt]
        -0.0661+0.390i&0.0857-0.467i&0.786\\[6pt]
        0.224-0.453i&-0.262+0.542i&0.594+0.170i\\[6pt]
        \end{pmatrix}.\label{lepton transformation2}
\end{align}

The best fit values of the leptonic parameters corresponding to the inverted mass ordering of neutrinos are found to be:
\begin{align}
\epsilon^\ell&=4.13\times10^{-5}-3.45\times10^{-4}i\,{\rm GeV},&&\quad \epsilon^\nu = -1.34-1.61i\,{\rm GeV},\nonumber \\
m_1^\ell&=-4.01\times10^{-3}-5.85\times10^{-2}i\,{\rm GeV},&&\quad m_1^\nu = -5.35\times10^{-2}-8.08\times10^{-3}i\,{\rm GeV},\nonumber\\
m_2^\ell&=9.83\times10^{-1}-1.79\times10^{-1}i\,{\rm GeV},&&\quad m_2^\nu = -1.05\times10^{-4}-1.31\times10^{-4}i\,{\rm GeV},\nonumber \\
m_3^\ell&=1.36\times10^{0}+4.44\times10^{-1}i\,{\rm GeV},&&\quad m_3^\nu =-1.55\times10^{0}+3.25\times10^{1}i\,{\rm GeV},\nonumber\\
m_4^\ell&=-2.09\times10^{-2}-1.08\times10^{-2}i\,{\rm GeV},&&\quad m_4^\nu = -5.20\times10^{-1}-3.10\times10^{-1}i\,{\rm GeV},\nonumber\\
M_1&=8.91\times10^{10}-7.27\times10^9i\,{\rm GeV},&&\quad M_3=8.16\times10^{16}+2.35\times10^{16}i\,{\rm GeV}.
\label{equ:lepton_inputs inverse}
\end{align}
The corresponding numerical values of the lepton mass matrices  $M^{\ell,\nu}$  and the mixing matrices $U_{\ell(L,R)}$ and $U_\nu$ are
\begin{align}
    \frac{M_e}{\mathrm{GeV}}=\begin{pmatrix}
    (0.413+3.446i)\times10^{-4}&(8.905-6.546i)\times10^{-3}&(1.278+1.384i)\times10^{-1}\\[6pt]
    (8.905-6.546i)\times10^{-3}&(0.0805+1.166i)\times10^{-1}&(9.827-1.791i)\times10^{-1} \\[6pt]
    (-0.388-4.418i)\times10^{-3}&(-2.089-1.082i)\times10^{-2}&(1.3634-0.444i)\times10^{0} \\[6pt]
    \end{pmatrix},\label{lepton mass matrix result inverse}
    \end{align}
    \begin{align}
    \frac{M_{\nu}^{\rm Majorana}}{10^{-12}\mathrm{GeV}}=\begin{pmatrix}
    -12.855+47.350i&0.465+0.0564i&2.426+1.245i\\[6pt]
    0.465+0.0564i&-14.993+42.862i&0.532+13.626i \\[6pt]
    2.426+1.245i&0.532+13.626i&1.733+3.649i \\[6pt]
    \end{pmatrix},\label{neutrino mass matrix result inverse}
    \end{align}
    \begin{align}
    U_{eL}=\begin{pmatrix}
    0.981&-0.0578-0.147i&0.0446+0.0971i\\[6pt]
    0.0150-0.0705i&0.819&0.565+0.0757i\\[6pt]
    -0.0576+0.167i&-0.547+0.0773i &0.815 \\[6pt]
    \end{pmatrix},\label{lepton transformation1 inverse}
    \end{align}
    \begin{align}
    U_{eR}=\begin{pmatrix}
    -0.155-0.985i&0.0677+0.0271i&0.00238+0.00454i\\[6pt]
    0.0680-0.0266i&0.165-0.983i&-0.00221-0.0314i\\[6pt]
    0.00413+0.000759i&-0.0271-0.0163i &0.950+0.309i \\[6pt]
    \end{pmatrix},
    \end{align}
    \begin{align}
        U_\nu=\begin{pmatrix}
        0.842&0.536+0.0291i&-0.0124+0.0496i\\[6pt]
        -0.521+0.0136i&0.803&-0.268+0.105\\[6pt]
        -0.136-0.00986i&0.230+0.117i&0.956\\[6pt]
        \end{pmatrix}.\label{lepton transformation2 inverse}
\end{align}

Although there is still no accurate measurement of the CP phase ($\delta_{CP}$) of the PMNS matrix, here we give the best-fit numerical prediction for its value from our model fits:
\begin{eqnarray}
    \delta_{CP}&=&119.95^\circ (\textbf{NO})
    \nonumber\\
    \delta_{CP}&=&121.48^\circ (\textbf{IO})
    \label{eq:CPnu}
\end{eqnarray}
Although these quoted values correspond to the best fit we obtained, variations from these predictions cannot exceed a few degrees.

Our lepton fit predicts the values of effective neutrino mass in Tritium beta decay ($m_\beta$)~\cite{Huang:2019tdh}, effective Majorana neutrino mass in neutrinoless double beta decay($m_{\beta\beta}$)~\cite{Bilenky:2014uka} and the sum of neutrino masses in cosmology($\Sigma$)~\cite{RoyChoudhury:2018gay}. The results are listed in~\autoref{table: neutrino mass} obtained by using the formulas~\cite{Formaggio:2021nfz,Fogli:2004as}:
\begin{align}
    \langle m_\beta \rangle \equiv \left\{\sum_{i=1} \mid U_{ei} \mid^2 m_i^2\right\}^{\frac{1}{2}}
\end{align}
\begin{align}
    \langle m_{\beta\beta} \rangle \equiv \mid\sum_{i=1} U_{ei}^2 m_i\mid
\end{align}
\begin{align}
    \Sigma \equiv \sum_{i=1} m_i
\end{align}

\begin{table}[!tb]
    \centering
\begin{tabular}{|c|c|c|}
\hline
Observables & Normal Ordering & Inverted Ordering \\
\hline
$m_\beta$ [eV]& 0.0116 & 0.0487 \\
\hline
$m_{\beta\beta}$[eV]&$0.0080$&$0.0479$\\
\hline
$\Sigma$ [eV]& 0.0697 & 0.0987  \\
\hline
\end{tabular}
\caption{Predictions of $m_\beta$,$m_{\beta\beta}$, $\Sigma$ in the Beta decay, Double Beta decay, and Cosmological measurements from normal ordering (NO) and inverted ordering (IO).}
\label{table: neutrino mass}
\end{table}

As we see, our predictions are well within the constraints $m_\beta<1.1$ eV~\cite{ParticleDataGroup:2022pth}, $m_{\beta\beta}<0.061-0.165$ eV~\cite{KamLAND-Zen:2016pfg}, and $\Sigma<0.13$ eV\cite{Planck:2018vyg}.

\subsection{Mass of lightest new Higgs boson}
\label{Higgs mass scan}

\begin{table}[!tb]
    \centering
\begin{tabular}{|c|c|c|c|}
\hline
Higgs & $h_2$ &$A_2 $& $H^{\pm}_2$ \\
\hline
$m$ [TeV]&17.667 &17.674 &17.671\\
\hline
Higgs&$h_3$&$A_3$&$H^{\pm}_3$\\
\hline
$m$ [TeV]&2402.170 &2402.170 &2402.170 \\
\hline
$\Delta m_K$ [GeV]& \multicolumn{3}{l|}{$3.36\times10^{-17}(\ll 3.48\times 10^{-15})$}\\
\hline
$\epsilon_K$ & \multicolumn{3}{l|}{$2.22\times10^{-3}(\lesssim 2.23\times10^{-3})$}\\
\hline
$\Delta m_{B_d}$ [GeV] & \multicolumn{3}{l|}{$9.32\times 10^{-17}(\ll 3.12\times10^{-13})$}\\
\hline
$\Delta m_{B_s}$ [GeV] & \multicolumn{3}{l|}{$6.99\times 10^{-14}(\ll 1.17\times10^{-11})$}\\
\hline
$\Delta m_D$ [GeV] & \multicolumn{3}{l|}{$6.24\times10^{-15}(\ll6.25\times 10^{-15})$}\\
\hline
$|d_n|$ [$e\cdot{\rm cm}$] &\multicolumn{3}{l|}{$1.71\times10^{-27}(\ll 1.20\times10^{-26})$}\\
\hline
\end{tabular}
\caption{The lightest heavy scalar masses satisfying the neutral meson mixing, neutron EDM as well as the quark and lepton sector measurements. The theoretical constraints (including BFB and unitarity) have been imposed. The corresponding neutral meson mass differences, the CP violation parameter for $K$ meson, and the neutron EDM are also listed.}
    \label{table:Higgs mass}
\end{table}

With the results from the above scan in the quark and lepton sector, the Yukawa couplings with the scalars are fixed up to the variations of the (complex) VEVs of the three scalar doublets. The VEVs and the scalar masses can be obtained from the Higgs potential parameters \{$v_3,\theta_2,\alpha_3,\alpha_4,\alpha_5,\beta_4,\beta_7,\lambda_2,\lambda_5,\lambda_6,\mu_5$\}. Note that, from the quark sector, we have already fixed the ratio of $v_1e^{i\theta_1}/v_2e^{i\theta_2}$. Further, we assume that the scalar $\eta_1$ in the Higgs basis in~\autoref{equ:Higgs_doublet_Higgs_basis} (and hence $h_1$ in~\autoref{equ:neutral_scalar_rotation}) is the observed 125 GeV Higgs. Hence we fixed the $M_{11}^2 = (125\ \rm GeV)^2$ and all other entries in the first row and column are fixed at zero. In this approximation, the lightest Higgs boson will behave just like the SM Higgs boson, and therefore constraints from Higgs precision observables are automatically satisfied.  This assumption provides 5 extra constraints on the Higgs potential parameters.
Then the Yukawa couplings in the mass basis are obtained according to~\autoref{equ:Yukawa_mass_basis}-\autoref{equ:rotation_combined}.
The neutral meson mixing and neutron EDM measurements as well as the theoretical constraints (BFB and perturbative unitarity) are imposed here to obtain the extra heavy scalar masses.\footnote{We still use the {\tt MultiNest}~\cite{Feroz:2008xx,Feroz:2013hea} and {\tt PyMultiNest}~\cite{Buchner:2014nha} to scan the parameter space. However, as we care more about the scalar masses, the likelihood is built using the heavy scalar masses. Any parameter point that violates the neutral meson mixing, neutron EDM measurements and the theoretical constraints is ignored.}
The parameters that provide the lightest heavy scalar mass are:
\begin{align}
& v_1=32.86\ \mathrm{GeV},\quad v_2=174.26\ \mathrm{GeV}, \quad v_3=170.50\ \mathrm{GeV}, \quad \theta_1=6.88\ \mathrm{rad}, \quad \theta_2=5.87\ \mathrm{rad},\nonumber\\
& \alpha_3=2.67\ \mathrm{rad}, \quad \alpha_4=2.25\ \mathrm{rad}, \quad \alpha_5=4.92\ \mathrm{rad}, \quad \beta_4=5.12\ \mathrm{rad}, \quad \beta_7=0.39\ \mathrm{rad},\nonumber\\
&\lambda_1=2.39,\quad \lambda_2=0.30, \quad \lambda_3=-0.30, \quad \lambda_4=2.85\times10^{-16}, \quad \lambda_5=0.91,\nonumber\\
&\lambda_6=-4.89, \quad \lambda_7=3.21\times10^{-16}, \quad \lambda_8=2.43, \quad \mu_5=3.53\times10^{5}\,{\rm GeV}.
\end{align}
The corresponding Higgs masses are listed in~\autoref{table:Higgs mass} from which we see that the lightest scalar mass which is needed to satisfy the neutral meson mixing, neutron EDM as well as the quark and lepton sector measurements is about 17 TeV.
This is beyond the reach of the current LHC but can be probed at the HE-LHC/FCC-hh~\cite{Cepeda:2019klc,FCC:2018byv} and/or high energy muon collider~\cite{Accettura:2023ked}. However, as mentioned above, CP violation measurement in the neutrino sector can be used to confirm or refute the model, see Eq. (\ref{eq:CPnu}).  The corresponding neutral meson mass differences, the CP violating parameter for $K$ meson and the neutron EDM are also listed in~\autoref{table:Higgs mass}. By comparing the theoretical predictions and the observations, it is clear that the constraints from $K$ meson (the CP violation parameter $\epsilon_K$) and $D$ meson  (the mass difference $\Delta m_D$) play important roles, while the neutron EDM is marginal.

\begin{figure}
    \centering
    \includegraphics[scale=0.5]{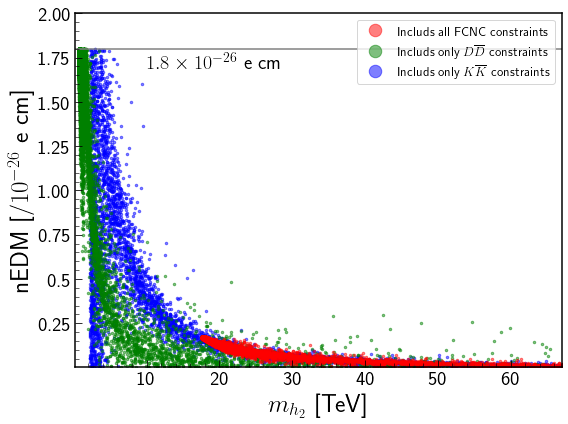}
    \caption{nEDM versus mass of new lightest Higgs boson under necessary BFB and unitarity conditions and different flavor constraints.}
    \label{fig:edm diagram}
\end{figure}

In~\autoref{fig:edm diagram}, we plot the neutron electric dipole moment (nEDM) versus mass of the lightest heavy scalar ($h_2$) under different FCNC constraints (all of the cases satisfy the necessary BFB conditions and the perturbative unitarity conditions of the potential). As can be seen from the figure, when only considering the $D$ meson or $K$ meson constraints, the neutron EDM will be the dominant constraint. In such cases, we can have light heavy scalars that can be probed from the current/future colliders. However, when we combine the constraints from $D$ meson and $K$ meson systems, the lower bound on the heavy scalar mass is pushed to about 17 TeV which is beyond the reach of the current high energy colliders. The combined constraints from the $D$ meson and $K$ meson systems supersede the nEDM constraint. Further, in each scan including only the $D$ meson or the $K$ meson constraints, the likelihood function in {\tt MultiNest} is set to prefer satisfying the corresponding constraints, the regions preferred by $D$ meson (green) or$K$ meson (blue) constraints do not overlap with each other too much. However, the region satisfying both constraints (red) overlaps with the region preferred by $K$ meson constraints. This implies when combining constraints from $D$ meson and $K$ meson system, those from $K$ meson system dominate.

\section{Conclusion}
\label{sec:conclusion}

In this paper, we have studied various phenomenological constraints on the 3HDM model with $S_3$ flavor symmetry.  The three fermion families and the three Higgs doublets transform as a doublet plus singlet under $S_3$.  Such an extension of the Higgs sector is well motivated, as it treats all fermion families and the Higgs families equivalently.
The $S_3$ symmetry is allowed to be broken softly in the scalar potential, which enables us to realize a decoupling limit. The Yukawa sector for quarks and leptons is constructed in a $S_3$-invariant way. Right-handed neutrinos, also transforming as a doublet plus singlet under $S_3$, are introduced to generate neutrino masses via the seesaw mechanism.   Numerica fits were carried to for the quark masses and the CKM mixing matrix, which fixes one of the ratios of VEVs, $v_1e^{i\theta_1}/v_2e^{i\theta_2}$. This information was fed into our fit to the lepton sector, and the resulting fits were found to be excellent.  In particular, the CP-violating phase in neutrino oscillations is determined to be $\delta_{CP}\simeq 120^0$, which holds for both the normal ordering and inverted ordering of neutrino masses.

We also investigated the lightest allowed mass of the new Higgs bosons consistent with FCNC processes that they mediate.  Not surprisingly, $K^0-\overline{K^0}$ mass splitting provides the most stringent constraint on the lightest boson mass, while $D^0-\overline{D^0}$ mixing also provides secondary constraints.  Furthermore, complex phases are needed in the Yukawa sector as well as the scalar potential in order to accommodate the CP phase in the CKM matrix (and in the PMNS matrix). In this context, CP violation mediated by the Higgs bosons is unavoidable and can lead to electric dipole moments of fermions. Hence, neutron EDM is considered as another constraint on the model. Within all these constraints, those from $K$ meson dominate, while those from the $D$ meson systems and neutron EDM are marginal, which are clearly shown in~\autoref{table:Higgs mass} and in~\autoref{fig:edm diagram}.

Under all these constraints, the lightest heavy scalar has to be heavier than 17 TeV, in order to have suppressed contributions to the neutral meson mixings. Such heavy masses for the new scalars are beyond the reach of the current collider experiments, but may be probed in future collider. Owing to the $S_3$ symmetry, two of the right-handed neutrinos are nearly degenerate, which may be a promising setup to generate lepton asymmetry resonantly in the early universe.
In summary, the 3HDM model with softly broken $S_3$ symmetry is able to explain simultaneously the fermion masses and mixings in quark and lepton sectors with a firm prediction on the leptonic CP phase.
Even when the masses of the heavy scalars exceed 17 TeV, their contributions to neutron EDM can be significant, which may be probed in the next generation of EDM experiments \cite{Alarcon:2022ero}.

\begin{acknowledgments}
The work of KSB was supported in part by the United States Department of Energy Grant No. DE-SC0016013. YW is partially supported by National Natural Science Foundation of China (NNSFC) under grant No. 12305112, and partially by China Ministry of Science and Technology under grant No. 2022YFA1605000. The computing for this project was performed at the High-Performance Computing Center at Oklahoma State University, supported in part through the National Science Foundation grant OAC-1531128.
\end{acknowledgments}

\appendix

\section{Minimization conditions}
\label{minimization}
Here we present the expressions for the Higgs potential minimization conditions leading to the VEVs, see~\autoref{min equation}:
\begin{align}
\mu_0^2= & -\lambda_1(v_1^2+v_2^2)+\lambda_2\cos\alpha_3\csc(\theta_1-\theta_2-\alpha_3)\sin(\theta_1-\theta_2)(v_1^2+v_2^2) \nonumber\\
& +\lambda_3\cos(\theta_1-\theta_2)\csc(\theta_1-\theta_2-\alpha_3)\sin \alpha_3(v_1^2+v_2^2)\nonumber\\
& +\lambda_4\frac{v_3}{8v_1^2v_2}\csc(\theta_1-\theta_2-\alpha_3)\csc(\theta_1+\alpha_4)\nonumber\\
&\times\Bigl(2\sin(\theta_1+\alpha_4)\bigl(-2\sin(\theta_1-\alpha_3+\beta_4)+\sin(\theta_1+\alpha_3+\beta_4)\bigl)v_1^4\nonumber\\
&+\bigl(\cos(2\theta_1-2\theta_2-\alpha_3+\alpha_4-\beta_4)-2\cos(\alpha_3+\alpha_4-\beta_4)\nonumber\\
& -\cos(2\theta_1-2\theta_2-\alpha_3-\alpha_4+\beta_4)+2\cos(\alpha_3-\alpha_4+\beta_4)\nonumber\\
&-\cos(2\theta_1+\alpha_3+\alpha_4+\beta_4)+\cos(2\theta_2+\alpha_3+\alpha_4+\beta_4)\bigl)v_1^2v_2^2\nonumber\\
&+2\sin(\theta_2+\alpha_3+\alpha_4)\sin(\theta_2+\beta_4)v_2^4\Bigl)-\frac{1}{2}(\lambda_5+\lambda_6)v_3^2\nonumber\\
&-\lambda_7\frac{v_3^2}{2v_1^2}\Bigl(\bigl(\cos(2\theta_1+\beta_7)+\cos(2\theta_2+\beta_7)-\cot(\theta_1+\alpha_4)\sin(2\theta_1+\beta_7)\nonumber \\
& +\cot(\theta_1-\theta_2-\alpha_3)\sin(2\theta_2+\beta_7)\bigl)v_1^2+\csc(\theta_1-\theta_2-\alpha_3)\csc(\theta_1+\alpha_4)\nonumber \\
& \times \sin(\theta_2+\alpha_3+\alpha_4)\sin(2\theta_2+\beta_7)v_2^2\Bigl)+\mu_5^2\frac{v_3}{4v_1^2v_2}\csc(-\theta_1+\theta_2+\alpha_3)\nonumber\\
&\times\bigl(\sin(\theta_1-\alpha_3+\alpha_5)v_1^2+\csc(\theta_1+\alpha_4)\sin(\theta_2+\alpha_3+\alpha_4)\nonumber\\
&\times\sin(\theta_2+\alpha_5)v_2^2\bigl),\label{mini1}\end{align}
\begin{align}
\mu_1^2= &\lambda_4\frac{v_2}{2v_3}\csc(\theta_1+\alpha_4)\Bigl(\bigl(2\sin(-\theta_1+\theta_2-\alpha_4+\beta_4)
+\sin(\theta_1-\theta_2-\alpha_4+\beta_4)\bigl)v_1^2\nonumber\\
& +\sin(\theta_1-\theta_2+\alpha_4-\beta_4)v_2^2\Bigl) -\frac{1}{2}(\lambda_5+\lambda_6)(v_1^2+v_2^2)\nonumber\\
&+\lambda_7\csc(\theta_1+\alpha_4)\bigl(\sin(\theta_1-\alpha_4+\beta_7)v_1^2-\sin(\theta_1-2\theta_2+\alpha_4-\beta_7)v_2^2\bigl)\nonumber\\
&-\lambda_8v_3^2-\mu_5^2\frac{v_2}{2v_3}\csc(\theta_1+\alpha_4)\sin(\theta_1-\theta_2+\alpha_4-\alpha_5),
\end{align}
\begin{align}
\mu_2^2= & (\lambda_2+\lambda_3)\cos\alpha_3\csc(\theta_1-\theta_2-\alpha_3)\sin(\theta_2-\theta_1)(v_1^2-v_2^2)\nonumber\\
& +\lambda_4\frac{v_3}{4v_1^2v_2}\Bigl(\csc(-\theta_1+\theta_2+\alpha_3)\bigl(2\sin(-\theta_1+\alpha_3-\beta_4)+\sin(\theta_1+\alpha_3+\beta_4)\bigl)v_1^4\nonumber\\
& +\bigl(-2\cos(2\theta_1-\theta_2+\beta_4)-7\cos(\theta_2+\beta_4)+\cot(\theta_1-\theta_2-\alpha_3)\nonumber\\
&\times\sin(2\theta_1-\theta_2+\beta_4)+\cot(\theta_1+\alpha_4)\sin(2\theta_1-\theta_2+\beta_4)\nonumber\\
&-3\cot(\theta_1-\theta_2-\alpha_3)\sin(\theta_2+\beta_4)+2\cot(\theta_1+\alpha_4)\sin(\theta_2+\beta_4)\bigl)v_1^2v_2^2\nonumber\\
& +\csc(\theta_1-\theta_2-\alpha_3)\csc(\theta_1+\alpha_4)\sin(\theta_2+\alpha_3+\alpha_4)\sin(\theta_2+\beta_4)v_2^4\Bigl)\nonumber\\
& -\lambda_7\frac{v_3^2}{2v_1^2}\Bigl(\bigl(\cos(2\theta_1+\beta_7)-\cos(2\theta_2+\beta_7)-\cot(\theta_1+\alpha_4)\sin(2\theta_1+\beta_7)\nonumber\\
& -\cot(\theta_1-\theta_2-\alpha_3)\sin(2\theta_2+\beta_7)\bigl)v_1^2+\csc(\theta_1-\theta_2-\alpha_3)\csc(\theta_1+\alpha_4)\nonumber\\
&\times\sin(\theta_2+\alpha_3+\beta_4)\sin(2\theta_2+\beta_7)v_2^2\Bigl)+\mu_5^2\frac{v_3}{4v_1^2v_2}\csc(\theta_1-\theta_2-\alpha_3)\nonumber\\
&\times\bigl(\sin(\theta_1-\alpha_3+\alpha_5)v_1^2-\csc(\theta_1+\alpha_4)\sin(\theta_2+\alpha_3+\alpha_4)\sin(\theta_2+\alpha_5)v_2^2\bigl),\end{align}
\begin{align}
\mu_3^2= &2(\lambda_2+\lambda_3)v_1v_2\csc(\alpha_3-\theta_1+\theta_2)\sin\bigl(2(\theta_1-\theta_2)\bigl)+\lambda_4\frac{v_3}{v_1}\csc(\alpha_3-\theta_1+\theta_2)\nonumber\\
&\times\Bigl(\bigl(\sin(\beta_4+2\theta_1-\theta_2)-2\sin(\beta_4+\theta_2)\bigl)v_1^2+\sin(\beta_4+\theta_2)v_2^2\Bigl)\nonumber\\
&+\lambda_7\frac{2v_2v_3^2}{v_1}\csc(\theta_1-\theta_2-\alpha_3)\sin(2\theta_2+\beta_7)+\mu_5^2\frac{v_3}{v_1}\csc(\theta_1-\theta_2-\alpha_3)\sin(\theta_2+\alpha_5),
\end{align}
\begin{align}
\mu_4^2=&\lambda_4\frac{v_2}{v_1}\csc(\theta_1+\alpha_4)\Bigl(-\bigl(\sin(2\theta_1-\theta_2+\beta_4)+2\sin(\theta_2+\beta_4)\bigl)v_1^2+\sin(\theta_2+\beta_4)v_2^2\Bigl)\nonumber\\
&-\lambda_7\frac{2v_3}{v_1}\csc(\theta_1+\alpha_4)\bigl(\sin(2\theta_1+\beta_7)v_1^2+\sin(2\theta_2+\beta_7)v_2^2\bigl)\nonumber\\
&-\mu_5^2\frac{v_2}{v_1}\csc(\theta_1+\alpha_4)\sin(\theta_2+\alpha_5).\label{mini5}
\end{align}
\section{Higgs Mass Matrices}
\subsection{Neutral Higgs Mass Matrix}
\label{Neutral Higgs Mass Matrix}
We present here the expressions for the nonzero elements of the $6\times6$  mass matrix for neutral Higgs boson mixing. Shorthand notations in~\autoref{equ:scalar_higgs_basis} are adopted. We use these expressions with formulas for neutral meson mixing~\autoref{FCNC equation} and nEDM~\autoref{equu:EDM_fermion} to scan for the lightest new Higgs boson that satisfies the various constraints we discussed in~\autoref{Higgs mass scan}.
\begin{align}
M^2_{11}= & \frac{2}{v^2}\biggl(\lambda_1v_{12}^4-4\lambda_2\sin^2(\theta_1-\theta_2)v_1^2v_2^2+\lambda_3\Bigl(v_1^4+2\cos\bigl(2(\theta_1-\theta_2)\bigl)v_1^2v_2^2+v_2^4\Bigl)\nonumber\\
& +2\lambda_4v_2v_3\Bigl(\bigl(\cos(2\theta_1-\theta_2+\beta_4)+2\cos(\theta_2+\beta_4)\bigl)v_1^2-\cos(\theta_2+\beta_4)v_2^2\Bigl)\nonumber\\
& +(\lambda_5+\lambda_6)v_{12}^2v_3^2+2\lambda_7v_3^2\bigl(\cos(2\theta_1+\beta_7)v_1^2+\cos(2\theta_2+\beta_7)v_2^2\bigl)+\lambda_8v_3^4\biggl),\label{D1}
\end{align}
\begin{align}
M^2_{12}=&\frac{2}{vv_{23}}\biggl(\lambda_1v_2v_{12}^2v_3-2\lambda_2v_1^2v_2v_3\sin^2(\theta_1-\theta_2)+\lambda_3v_2v_3\Bigl(\cos\bigl(2(\theta_1-\theta_2)\bigl)v_1^2+v_2^2\Bigl)\nonumber\\
&+\frac{1}{2}\lambda_4\Bigl(\cos(\theta_2+\beta_4)v_2^2(v_2^2-3v_3^2)-\bigl(\cos(2\theta_1-\theta_2+\beta_4)\nonumber\\
&+2\cos(\theta_2+\beta_4)\bigl)v_1^2(v_2^2-v_3^2)\Bigl)
 -\frac{1}{2}(\lambda_5+\lambda_6)v_2v_3(v_1^2+v_2^2-v_3^2)\nonumber\\
&+\lambda_7v_2v_3\bigl(-\cos(2\theta_1+\beta_7)v_1^2
+\cos(2\theta_2+\beta_7)(-v_2^2+v_3^2)\bigl)-\lambda_8v_2v_3^3\biggl),\label{D2}
\end{align}
\begin{align}
M^2_{13}=& \frac{2}{v^2v_{23}}\biggl(-\lambda_1v_1v_3^2(v_1^2+v_2^2)+2\lambda_2v_1v_2^2\sin^2(\theta_1-\theta_2)(-v_1^2+v_2^2+v_3^2) \nonumber\\
&-\lambda_3v_1\Bigl(-2\sin^2(\theta_1-\theta_2)v_2^4
+\cos\bigl(2(\theta_1-\theta_2)\bigl)v_2^2v_3^2+v_1^2\bigl(2\sin^2(\theta_1-\theta_2)v_2^2+v_3^2\bigl)\Bigl)\nonumber\\
&+\lambda_4v_1v_2v_3\Bigl(\bigl(\cos(2\theta_1-\theta_2+\beta_4)+2\cos(\theta_2+\beta_4)\bigl)v_1^2-\bigl(\cos(2\theta_1-\theta_2+\beta_4)\nonumber\\
&+4\cos(\theta_2+\beta_4)\bigl)v_2^2-\bigl(\cos(2\theta_1-\theta_2+\beta_4)+2\cos(\theta_2+\beta_4)\bigl)v_3^2\Bigl)\nonumber\\
&+\frac{1}{2}(\lambda_5+\lambda_6)v_1v_3^2(v_1^2+v_2^2-v_3^2)
+\lambda_7v_1v_3^2\Bigl(\cos(2\theta_1+\beta_7)v_1^2\nonumber\\
&-\bigl(\cos(2\theta_1+\beta_7)-2\cos(2\theta_2+\beta_7)\bigl)v_2^2
-\cos(2\theta_1+\beta_7)v_3^2\Bigl)+\lambda_8v_1v_3^4\biggl),
\end{align}
\begin{align}
M^2_{15}=&\frac{2}{vv_{23}}\Biggl((\lambda_2+\lambda_3)v_1^2v_2v_3\sin\bigl(2(\theta_1-\theta_2)\bigl)+\frac{1}{2}\lambda_4\biggl(\sin(\theta_2+\beta_4)v_2^2v_{23}^2\nonumber\\
&-v_1^2\Bigl(\bigl(\sin(2\theta_1-\theta_2+\beta_4)+2\sin(\theta_2+\beta_4)\bigl)v_2^2+\bigl(\sin(-2\theta_1+\theta_2-\beta_4)\nonumber\\
&+2\sin(\theta_2+\beta_4)\bigl)v_3^2\Bigl)\biggl)-\lambda_7v_2v_3\bigl(\sin(2\theta_1+\beta_7)v_1^2+\sin(2\theta_2+\beta_7)v_{23}^2\bigl)\Biggl),
\end{align}
\begin{align}
M^2_{16}=\frac{2v_1}{v_{23}}\Bigl((\lambda_2+\lambda_3)\sin\bigl(2(\theta_1-\theta_2)\bigl)v_2^2+\lambda_4\sin(2\theta_1-\theta_2+\beta_4)v_2v_3+\lambda_7\sin(2\theta_1+\beta_7)v_3^2\Bigl),\label{D5}
\end{align}
\begin{align}
M^2_{22}=&\frac{1}{v_{23}^2}\Biggl(2(\lambda_1-\lambda_5-\lambda_6)v_2^2v_3^2+v_3^2\Bigl(2\lambda_3v_2^2+(\lambda_2+\lambda_3)\cot(\theta_1-\theta_2-\alpha_3)\sin\bigl(2(\theta_1-\theta_2)\bigl)v_1^2\Bigl)\nonumber\\
&+\frac{\lambda_4}{2v_2v_3}\biggl(\csc(\theta_1+\alpha_4)\sin(\theta_1-\theta_2+\alpha_4-\beta_4)v_2^6+6\cos(\theta_2+\beta_4)v_2^4v_3^2\nonumber\\
&-\csc(\theta_1-\theta_2-\alpha_3)\bigl(2\sin(\theta_1-2\theta_2-\alpha_3-\beta_4)+\sin(\theta_1-\alpha_3+\beta_4)\bigl)v_2^2v_3^4+\nonumber\\
&v_1^2\Bigl(-\csc(\theta_1+\alpha_4)\bigl(2\sin(\theta_1-\theta_2+\alpha_4-\beta_4)-\sin(\theta_1-\theta_2-\alpha_4+\beta_4)\bigl)v_2^4\nonumber\\
&-2\bigl(\cos(2\theta_1-\theta_2+\beta_4)+2\cos(\theta_2+\beta_4)\bigl)v_2^2v_3^2+\csc(\theta_1-\theta_2-\alpha_3)\nonumber\\
&\times\bigl(-2\sin(\theta_1-\alpha_3+\beta_4)+\sin(\theta_1+\alpha_3+\beta_4)\bigl)v_3^4\Bigl)\biggl)\nonumber\\
&+\lambda_7\bigl(\cot(\theta_1+\alpha_4)\sin(2\theta_1+\beta_7)v_1^2v_2^2+\cot(\theta_1+\alpha_4)\sin(2\theta_2+\beta_7)v_2^4\nonumber\\
&-4\cos(2\theta_2+\beta_7)v_2^2v_3^2-\cot(\theta_1-\theta_2-\alpha_3)\sin(2\theta_2+\beta_7)v_3^4\bigl)+2\lambda_8v_2^2v_3^2\nonumber\\
&-\frac{\mu_5^2}{2v_2v_3}\bigl(\csc(\theta_1+\alpha_4)\sin(\theta_1-\theta_2+\alpha_4-\alpha_5)v_2^4+2\cos(\theta_2+\alpha_5)v_2^2v_3^2\nonumber\\
&+\csc(\theta_1-\theta_2-\alpha_3)\sin(\theta_1-\alpha_3+\alpha_5)v_3^4\bigl)\Biggl),
\end{align}
\begin{align}
M^2_{23}= &\frac{2}{v_{23}^2v}\Biggl((-\lambda_1+\lambda_5+\lambda_6-\lambda_8)v_1v_2v_3^3\nonumber\\
&+\frac{1}{2}\lambda_2\csc(\theta_1-\theta_2-\alpha_3)\sin(\theta_1-\theta_2)v_1v_2v_3\Bigl(\bigl(\cos(2\theta_1-2\theta_2-\alpha_3)+\cos\alpha_3\bigl)v_1^2\nonumber\\
&-\bigl(\cos(2\theta_1-2\theta_2-\alpha_3)-3\cos \alpha_3\bigl)v_{23}^2\Bigl)\nonumber\\
&-\frac{1}{4}\lambda_3\csc(\theta_1-\theta_2-\alpha_3)v_1v_2v_3\Bigl(-2\cos(\theta_1-\theta_2-\alpha_3)\sin\bigl(2(\theta_1-\theta_2)\bigl)v_1^2\nonumber\\
&+2\bigl(\cos(2\theta_1-2\theta_2-\alpha_3)-3\cos \alpha_3\bigl)\sin(\theta_1-\theta_2)v_2^2+\bigl(\sin(3\theta_1-3\theta_2-\alpha_3)\nonumber\\
&-3\sin(\theta_1-\theta_2+\alpha_3)\bigl)v_3^2\Bigl)
-\frac{\lambda_4}{16v_1}\Biggl(-4\sin(\theta_2+\beta_4)v_2^2v_{23}^2\bigl(\cot(\theta_1+\alpha_4)v_2^2\nonumber\\
&+\cot(\theta_1-\theta_2-\alpha_3)v_3^2\bigl)+4v_1^4\Bigl(\cot(\theta_1+\alpha_4)\bigl(\sin(2\theta_1-\theta_2+\beta_4)+2\sin(\theta_2+\beta_4)\bigl)v_2^2\nonumber\\
&+\cot(\theta_1-\theta_2-\alpha_3)\bigl(-\sin(2\theta_1-\theta_2+\beta_4)+2\sin(\theta_2+\beta_4)\bigl)v_3^2\Bigl)\nonumber\\
&+v_1^2\Bigl(-2\csc(\theta_1+\alpha_4)\bigl(7\sin(\theta_1-\theta_2+\alpha_4-\beta_4)-3\sin(\theta_1-\theta_2-\alpha_4+\beta_4)\nonumber\\
&+\sin(3\theta_1-\theta_2+\alpha_4+\beta_4)+5\sin(\theta_1+\theta_2+\alpha_4+\beta_4)\bigl)v_2^4\nonumber\\
&+\bigl(-3\cos(2\theta_1-2\theta_2-\alpha_3+\alpha_4-\beta_4)+5\cos(\alpha_3+\alpha_4-\beta_4)\nonumber\\
&-2\cos(2\theta_1-2\theta_2-\alpha_3-\alpha_4+\beta_4)-9\cos(2\theta_1-\alpha_3+\alpha_4+\beta_4)\nonumber\\
&+2\cos(2\theta_1+\alpha_3+\alpha_4+\beta_4)+7\cos(2\theta_2+\alpha_3+\alpha_4+\beta_4)\bigl)\csc(\theta_1-\theta_2-\alpha_3)\nonumber\\
&\times\csc(\theta_1+\alpha_4)v_2^2v_3^2+2\csc(\theta_1-\theta_2-\alpha_3)\bigl(2\sin(\theta_1-2\theta_2-\alpha_3-\beta_4)\nonumber\\
&+6\sin(\theta_1-\alpha_3+\beta_4)\nonumber+\sin(3\theta_1-2\theta_2-\alpha_3+\beta_4)-3\sin(\theta_1+\alpha_3+\beta_4)\bigl)v_3^4\Bigl)\biggl)\nonumber\\
&-\lambda_7\frac{v_2v_3}{8v_1}\biggl(4\cot(\theta_1+\alpha_4)\sin(2\theta_1+\beta_7)v_1^4+4\sin(2\theta_2+\beta_7)v_{23}^2\bigl(\cot(\theta_1+\alpha_4)v_2^2\nonumber\\
&+\cot(\theta_1-\theta_2-\alpha_3)v_3^2\bigl)+\csc(\theta_1+\alpha_4)v_1^2\Bigl(-4(\cos(2\theta_1-\theta_2+\alpha_4)-3\cos(\theta_2+\alpha_4))\nonumber\\
&\times\sin(\theta_1+\theta_2+\beta_7)v_2^2+\bigl(3\cos(2\theta_1-3\theta_2-\alpha_3+\alpha_4-\beta_7)\nonumber\\
&-3\cos(2\theta_1-\theta_2-\alpha_3-\alpha_4+\beta_7)-\cos(\theta_2-\alpha_3-\alpha_4+\beta_7)\nonumber\\
&+3\cos(\theta_2+\alpha_3-\alpha_4+\beta_7)+\cos(4\theta_1-\theta_2-\alpha_3+\alpha_4+\beta_7)+\cos(2\theta_1+\theta_2-\alpha_3+\alpha_4+\beta_7)\nonumber\\
&-\cos(2\theta_1+\theta_2+\alpha_3+\alpha_4+\beta_7)-3\cos(3\theta_2+\alpha_3+\alpha_4+\beta_7)\bigl)\csc(\theta_1-\theta_2-\alpha_3)v_3^2\Bigl)\biggl)\nonumber\\
&-\mu_5^2\frac{v^2}{4v_1}\sin(\theta_2+\alpha_5)\bigl(\cot(\theta_1+\alpha_4)v_2^2+\cot(\theta_1-\theta_2-\alpha_3)v_3^2\bigl)\Biggl),
\end{align}
\begin{align}
M^2_{25}=&\frac{1}{v_{23}^2}\biggl((\lambda_2+\lambda_3)v_1^2v_3^2\sin\bigl(2(\theta_1-\theta_2)\bigl)+\lambda_4v_2v_3\bigl(v_{23}^2\sin(\theta_2+\beta_4)
-\sin(2\theta_1-\theta_2+\beta_4)v_1^2\bigl)\nonumber\\
&+\lambda_7\bigl(\sin(2\theta_1+\beta_7)v_1^2v_2^2
+\sin(2\theta_2+\beta_7)(v_2^4-v_3^4)\bigl)\biggl),
\end{align}
\begin{align}
M^2_{26}=&\frac{1}{v_{23}^2}\Biggl((\lambda_2+\lambda_3)v_1v_2v_3v\sin\bigl(2(\theta_1-\theta_2)\bigl)+\lambda_4\frac{v}{2v_1}\biggl(-\sin(\theta_2+\beta_4)v_2^2v_{23}^2\nonumber\\
&+v_1^2\Bigl(\bigl(-\sin(2\theta_1-\theta_2+\beta_4)+2\sin(\theta_2+\beta_4)\bigl)v_2^2+\bigl(\sin(2\theta_1-\theta_2+\beta_4)\nonumber\\
&+2\sin(\theta_2+\beta_4)\bigl)v_3^2\Bigl)\biggl)+\lambda_7\frac{v_2v_3v}{v_1}\bigl(-\sin(2\theta_1+\beta_7)v_1^2\nonumber\\
&+\sin(2\theta_2+\beta_7)v_{23}^2\bigl)+\mu_5^2\frac{vv_{23}^2}{2v_1}\sin(\theta_2+\alpha_5)\Biggl),
\end{align}
\begin{align}
M^2_{33}=&\frac{1}{v_{23}^2v^2}\Biggl(2(\lambda_1-\lambda_5-\lambda_6+\lambda_8)v_1^2v_3^4\nonumber\\
&+\lambda_2\csc(\theta_1-\theta_2-\alpha_3)\sin(\theta_1-\theta_2)v_2^2\Bigl(\bigl(\cos(2\theta_1-2\theta_2-\alpha_3)+\cos\alpha_3\bigl)v_1^4\nonumber\\
&-2\bigl(\cos(2\theta_1-2\theta_2-\alpha_3)-3\cos\alpha_3\bigl)v_1^2v_{23}^2+\bigl(\cos(2\theta_1-2\theta_2-\alpha_3)+\cos\alpha_3\bigl)v_{23}^4\Bigl)\nonumber\\
&+\lambda_3\biggl(\cot(\theta_1-\theta_2-\alpha_3)\sin\bigl(2(\theta_1-\theta_2)\bigl)v_1^4v_2^2+\cot(\theta_1-\theta_2-\alpha_3)\sin\bigl(2(\theta_1-\theta_2)\bigl)v_2^2v_{23}^4\nonumber\\
&+v_1^2\Bigl(-2\bigl(\cos(2\theta_1-2\theta_2-\alpha_3)-3\cos\alpha_3\bigl)\csc(\theta_1-\theta_2-\alpha_3)\sin(\theta_1-\theta_2)v_2^4\nonumber\\
&-2\bigl(\cos(2\theta_1-2\theta_2-\alpha_3)\nonumber-3\cos\alpha_3\bigl)\csc(\theta_1-\theta_2-\alpha_3)\sin(\theta_1-\theta_2)v_2^2v_3^2+2v_3^4\Bigl)\biggl)\nonumber\\
&+\frac{1}{4}\lambda_4v_2v_3\Bigl(4\bigl(\cos(2\theta_1-\theta_2+\beta_4)+2\cos(\theta_2+\beta_4)\bigl)v_1^4+2\csc(\theta_1-\theta_2-\alpha_3)\nonumber\\
&\times\bigl(2\sin(-\theta_1+\alpha_3-\beta_4)+\sin(\theta_1+\alpha_3+\beta_4)\bigl)v_1^4+2\csc(\theta_1+\alpha_4)\nonumber\\
&\times\bigl(2\sin(-\theta_1+\theta_2-\alpha_4+\beta_4)+\sin(\theta_1-\theta_2-\alpha_4+\beta_4)\bigl)v_1^4-8\bigl(2\cos(2\theta_1-\theta_2+\beta_4)\nonumber\\
&+7\cos(\theta_2+\beta_4)\bigl)v_1^2v_2^2+2\csc(\theta_1+\alpha_4)\sin(\theta_1-\theta_2+\alpha_4-\beta_4)v_1^2v_2^2\nonumber\\
&+2\csc(\theta_1-\theta_2-\alpha_3)\bigl(\sin(\theta_1-2\theta_2-\alpha_3-\beta_4)+2\sin(\theta_1-\alpha_3+\beta_4)\bigl)v_1^2v_2^2\nonumber\\
&-16\bigl(\cos(2\theta_1-\theta_2+\beta_4)+2\cos(\theta_2+\beta_4)\bigl)v_1^2v_3^2+2\bigl(-2\cos(\alpha_3+\alpha_4-\beta_4)+\cos(\alpha_3-\alpha_4+\beta_4)\nonumber\\
&-\cos(4\theta_1-2\theta_2-\alpha_3+\alpha_4+\beta_4)
+2\cos(2\theta_2+\alpha_3+\alpha_4+\beta_4)\bigl)\csc(\theta_1-\theta_2-\alpha_3)\nonumber\\
&\times\csc(\theta_1+\alpha_4)v_1^2v_{23}^2+4\csc(\theta_1-\theta_2-\alpha_3)\csc(\theta_1+\alpha_4)\sin(\theta_2+\alpha_3+\alpha_4)\nonumber\\
&\times\sin(\theta_2+\beta_4)v_2^2v_{23}^2+\bigl(-2\cos(\alpha_3+\alpha_4-\beta_4)+\cos(\alpha_3-\alpha_4+\beta_4)+2\cos(2\theta_2+\alpha_3\nonumber\\
&+\alpha_4+\beta_4)-\cos(4\theta_1-2\theta_2-\alpha_3+\alpha_4+\beta_4)\bigl)\csc(\theta_1-\theta_2-\alpha_3)\csc(\theta_1+\alpha_4)v_{23}^4\nonumber\\
&+\frac{2v_2^2v_{23}^4}{v_1^2}\csc(\theta_1-\theta_2-\alpha_3)\csc(\theta_1+\alpha_4)\sin(\theta_2+\alpha_3+\alpha_4)\sin(\theta_2+\beta_4)\Bigl)\nonumber\\
&+\lambda_7\frac{v_3^2}{v_1^2}\Biggl(\cot(\theta_1+\alpha_4)\sin(2\theta_1+\beta_7)v_1^6-\csc(\theta_1-\theta_2-\alpha_3)\nonumber\\
&\times\csc(\theta_1+\alpha_4)\sin(\theta_2+\alpha_3+\alpha_4)\sin(2\theta_2+\beta_7)v_2^2v_{23}^4+v_1^2v_{23}^2\Bigl(\bigl(\cot(\theta_1+\alpha_4)\nonumber\\
&\times\sin(2\theta_1+\beta_7)-2\csc(\theta_1-\theta_2-\alpha_3)\csc(\theta_1+\alpha_4)\sin(\theta_2+\alpha_3+\alpha_4)\nonumber\\
& \times \sin(2\theta_2+\beta_7)\bigl)v_2^2+\cot(\theta_1+\alpha_4)\sin(2\theta_1+\beta_7)v_3^2\Bigl)+v_1^4\biggl(\Bigl(-\csc(\theta_1+\alpha_4)\nonumber\\
&\times\sin(\theta_1-2\theta_2+\alpha_4-\beta_7)+2\cot(\theta_1+\alpha_4)\sin(2\theta_1+\beta_7)+\csc(\theta_1-\theta_2-\alpha_3)\nonumber\\
&\times\bigl(3\sin(\theta_1-3\theta_2-\alpha_3-\beta_7)+4\cos\alpha_3\sin(\theta_1+\theta_2+\beta_7)-2\sin(3\theta_1-\theta_2\nonumber\\
&-\alpha_3+\beta_7)\bigl)\Bigl)v_2^2+2\bigl(-2\cos(2\theta_1+\beta_7)+\cot(\theta_1+\alpha_4)\sin(2\theta_1+\beta_7)\bigl)v_3^2\biggl)\Biggl)\nonumber\\
&-\mu_5^2\frac{v_2v_3v^4}{2v_1^2}\csc(\theta_1-\theta_2-\alpha_3)\csc(\theta_1+\alpha_4)\sin(\theta_2+\alpha_3+\alpha_4)\sin(\theta_2+\alpha_5)\Biggl),
\end{align}
\begin{align}
M^2_{35}=&\frac{1}{v_{23}^2v}\Biggl((\lambda_2+\lambda_3)\sin\bigl(2(\theta_1-\theta_2)\bigl)v_1v_2v_3(v_1^2-v_2^2-v_3^2)\nonumber\\
&+\frac{\lambda_4}{2v_1}\biggl(\sin(\theta_2+\beta_4)v_2^2v_{23}^4-v_1^4\Bigl(\bigl(\sin(2\theta_1-\theta_2+\beta_4)+2\sin(\theta_2+\beta_4)\bigl)v_2^2\nonumber\\
&+\bigl(\sin(-2\theta_1+\theta_2-\beta_4)+2\sin(\theta_2+\beta_4)\bigl)v_3^2\Bigl)+v_1^2v_{23}^2\Bigl(\bigl(\sin(2\theta_1-\theta_2+\beta_4)\nonumber\\
&+5\sin(\theta_2+\beta_4)\bigl)v_2^2+\bigl(\sin(-2\theta_1+\theta_2-\beta_4)+2\sin(\theta_2+\beta_4)\bigl)v_3^2\Bigl)\biggl)\nonumber\\
&-\lambda_7\frac{v_2v_3}{v_1}\Bigl(\sin(2\theta_1+\beta_7)v_1^4-\bigl(\sin(2\theta_1+\beta_7)-3\sin(2\theta_2+\beta_7)\bigl)v_1^2v_{23}^2\nonumber\\
&+\sin(2\theta_2+\beta_7)v_{23}^4\Bigl)-\mu_5^2\frac{v_{23}^2v^2}{2v_1}\sin(\theta_2+\alpha_5)\Biggl),
\end{align}
\begin{align}
M^2_{36}=&\frac{2v_1^2-v^2}{v_{23}^2}\Bigl((\lambda_2+\lambda_3)\sin\bigl(2(\theta_1-\theta_2)\bigl)v_2^2+\lambda_4v_2v_3\sin(2\theta_1-\theta_2+\beta_4)\nonumber\\
&+\lambda_7\sin(2\theta_1+\beta_7)v_3^2\Bigl),
\end{align}
\begin{align}
M^2_{55}=&\frac{1}{2v_{23}^2}\Biggl(-(\lambda_2+\lambda_3)\csc(\theta_1-\theta_2-\alpha_3)\bigl(\sin(3\theta_1-3\theta_2-\alpha_3)-3\sin(\theta_1-\theta_2+\alpha_3)\bigl)v_1^2v_3^2\nonumber\\
&+\frac{\lambda_4}{v_2v_3}\biggl(\csc(\theta_1+\alpha_4)\sin(\theta_1-\theta_2+\alpha_4-\beta_4)v_2^6+2\cos(\theta_2+\beta_4)v_2^4v_3^2\nonumber\\
&+\csc(\theta_1-\theta_2-\alpha_3)\sin(\theta_1-\alpha_3+\beta_4)v_2^2v_3^4+v_1^2\Bigl(-\csc(\theta_1+\alpha_4)\nonumber\\
&\times\bigl(2\sin(\theta_1-\theta_2+\alpha_4-\beta_4)-\sin(\theta_1-\theta_2-\alpha_4+\beta_4)\bigl)v_2^4\nonumber\\
&+2\bigl(\cos(2\theta_1-\theta_2+\beta_4)-2\cos(\theta_2+\beta_4)\bigl)v_2^2v_3^2\nonumber\\
&+\csc(\theta_1-\theta_2-\alpha_3)\bigl(-2\sin(\theta_1-\alpha_3+\beta_4)+\sin(\theta_1+\alpha_3+\beta_4)\bigl)v_3^4\Bigl)\biggl)\nonumber\\
&-\lambda_7\Bigl(\csc(\theta_1+\alpha_4)\bigl(-3\sin(\theta_1-\alpha_4+\beta_7)+\sin(3\theta_1+\alpha_4+\beta_7)\bigl)v_1^2v_2^2\nonumber\\
&+\csc(\theta_1+\alpha_4)\bigl(3\sin(\theta_1-2\theta_2+\alpha_4-\beta_7)+\sin(\theta_1+2\theta_2+\alpha_4+\beta_7)\bigl)v_2^4\nonumber\\
&+8\cos(2\theta_2+\beta_7)v_2^2v_3^2+\csc(\theta_1-\theta_2-\alpha_3)\bigl(\sin(\theta_1-3\theta_2-\alpha_3-\beta_7)\nonumber\\
&+3\sin(\theta_1+\theta_2-\alpha_3+\beta_7)\bigl)v_3^4\Bigl)-\frac{\mu_5^2}{v_2v_3}\bigl(\csc(\theta_1+\alpha_4)\sin(\theta_1-\theta_2+\alpha_4-\alpha_5)v_2^4\nonumber\\
&+2\cos(\theta_2+\alpha_5)v_2^2v_3^2+\csc(\theta_1-\theta_2-\alpha_3)\sin(\theta_1-\alpha_3+\alpha_5)v_3^4\bigl)\Biggl),
\end{align}
\begin{align}
M^2_{56}=&\frac{1}{2v_{23}^2}\Biggl(-(\lambda_2+\lambda_3)v_1v_2v_3v\bigl(\sin(3\theta_1-3\theta_2-\alpha_3)-3\sin(\theta_1-\theta_2+\alpha_3)\bigl)\nonumber\\
&\times\csc(\theta_1-\theta_2-\alpha_3)+\frac{\lambda_4v}{2v_1}\biggl(2\sin(\theta_2+\beta_4)v_2^2\bigl(\cot(\theta_1+\alpha_4)v_2^2\nonumber\\
&+\cot(\theta_1-\theta_2-\alpha_3)v_3^2\bigl)+v_1^2\Bigl(\csc(\theta_1+\alpha_4)\bigl(2\sin(\theta_1-\theta_2+\alpha_4-\beta_4)\nonumber\\
&-3\sin(\theta_1-\theta_2-\alpha_4+\beta_4)+\sin(3\theta_1-\theta_2+\alpha_4+\beta_4)\nonumber\\
&-2\sin(\theta_1+\theta_2+\alpha_4+\beta_4)\bigl)v_2^2+\csc(\theta_1-\theta_2-\alpha_3)\bigl(2\sin(\theta_1-2\theta_2-\alpha_3-\beta_4)\nonumber\\
&-2\sin(\theta_1-\alpha_3+\beta_4)-\sin(3\theta_1-2\theta_2-\alpha_3+\beta_4)+3\sin(\theta_1+\alpha_3+\beta_4)\bigl)v_3^2\Bigl)\biggl)\nonumber\\
&-\lambda_7\frac{v_2v_3v}{v_1}\Bigl(-\csc(\theta_1+\alpha_4)\bigl(-3\sin(\theta_1-\alpha_4+\beta_7)+\sin(3\theta_1+\alpha_4+\beta_7)\bigl)v_1^2\nonumber\\
&+2\sin(2\theta_2+\beta_7)\bigl(\cot(\theta_1+\alpha_4)v_2^2+\cot(\theta_1-\theta_2-\alpha_3)v_3^2\bigl)\Bigl)\nonumber\\
&-\mu_5^2\frac{v}{v_1}\sin(\theta_2+\alpha_5)\bigl(\cot(\theta_1+\alpha_4)v_2^2+\cot(\theta_1-\theta_2-\alpha_3)v_3^2\bigl)\Biggl),
\end{align}
\begin{align}
M^2_{66}=&\frac{1}{2v_{23}^2}\biggl(-(\lambda_2+\lambda_3)\csc(\theta_1-\theta_2-\alpha_3)\bigl(\sin(3\theta_1-3\theta_2-\alpha_3)-3\sin(\theta_1-\theta_2+\alpha_3)\bigl)v_2^2v^2\nonumber\\
&+\lambda_4\frac{v_2v_3v^2}{2v_1^2}\csc(\theta_1-\theta_2-\alpha_3)\csc(\theta_1+\alpha_4)\Bigl(\bigl(-2\cos(\alpha_3+\alpha_4-\beta_4)\nonumber\\
&-2\cos(2\theta_1-2\theta_2-\alpha_3-\alpha_4+\beta_4)+3\cos(\alpha_3-\alpha_4+\beta_4)\nonumber\\
&+\cos(4\theta_1-2\theta_2-\alpha_3+\alpha_4+\beta_4)-2\cos(2\theta_1+\alpha_3+\alpha_4+\beta_4)\nonumber\\
&+2\cos(2\theta_2+\alpha_3+\alpha_4+\beta_4)\bigl)v_1^2
+2\sin(\theta_2+\alpha_3+\alpha_4)\sin(\theta_2+\beta_4)v_2^2\Bigl)\nonumber\\
&+\lambda_7\frac{v^2v_3^2}{v_1^2}\csc(\theta_1+\alpha_4)\Bigl(\bigl(3\sin(\theta_1-\alpha_4+\beta_7)-\sin(3\theta_1+\alpha_4+\beta_7)\bigl)v_1^2\nonumber\\
&-2\csc(\theta_1-\theta_2-\alpha_3)\sin(\theta_2+\alpha_3+\alpha_4)\sin(2\theta_2+\beta_7)v_2^2\Bigl)\nonumber\\
&-\mu_5^2\frac{v^2v_2v_3}{v_1^2}\csc(\theta_1-\theta_2-\alpha_3)\csc(\theta_1+\alpha_4)\sin(\theta_2+\alpha_3+\alpha_4)\sin(\theta_2+\alpha_5)\biggl).\label{D15}
\end{align}
\subsection{Charged Higgs Mass Matrix}
\label{Charged Higgs Mass Matrix}
Here we present expressions for the nonzero elements of the $3\times3$ mass matrix for charged Higgs bosons. Shorthand notations in~\autoref{equ:scalar_higgs_basis} are adopted. We substitute these expressions to nEDM formula~\autoref{equu:EDM_fermion} to scan for the lightest new Higgs mass.
\begin{align}
M^{2}_{\pm22}=&\frac{1}{2v_2v_3v_{23}^2}\Biggl(4v_1^2v_2v_3^3\csc(\theta_1-\theta_2-\alpha_3)\bigl(\lambda_2\cos\alpha_3\sin(\theta_1-\theta_2)\nonumber\\
&+\lambda_3\sin\alpha_3\cos(\theta_1-\theta_2)\bigl)+\lambda_4\biggl(\csc(\theta_1+\alpha_4)\sin(\theta_1-\theta_2+\alpha_4-\beta_4)v_2^6\nonumber\\
&+2\cos(\theta_2+\beta_4)v_2^4v_3^2+\csc(\theta_1-\theta_2-\alpha_3)\sin(\theta_1-\alpha_3+\beta_4)v_2^2v_3^4\nonumber\\
&+v_1^2\Bigl(-\csc(\theta_1+\alpha_4)\bigl(2\sin(\theta_1-\theta_2+\alpha_4-\beta_4)-\sin(\theta_1-\theta_2-\alpha_4+\beta_4)   \bigl)v_2^4\nonumber\\
&-2\cos(\theta_2+\beta_4)v_2^2v_3^2+\csc(\theta_1-\theta_2-\alpha_3)\bigl(-2\sin(\theta_1-\alpha_3+\beta_4)\nonumber\\
&+\sin(\theta_1+\alpha_3+\beta_4)\bigl)v_3^4\Bigl)\biggl)-\lambda_6v_2v_3(v_1^2v_2^2+v_{23}^4)\nonumber\\
&-2\lambda_7v_2v_3\bigl(-\csc(\theta_1+\alpha_4)\sin(\theta_1-\alpha_4+\beta_7)v_1^2v_2^2\nonumber\\
&+\csc(\theta_1+\alpha_4)\sin(\theta_1-2\theta_2+\alpha_4-\beta_7)v_2^4+2\cos(2\theta_2+\beta_7)v_2^2v_3^2\nonumber\\
&+\csc(\theta_1-\theta_2-\alpha_3)\sin(\theta_1+\theta_2-\alpha_3+\beta_7)v_3^4\bigl)\nonumber\\
&-\mu_5^2\bigl(\csc(\theta_1+\alpha_4)\sin(\theta_1-\theta_2+\alpha_4-\alpha_5)v_2^4+2\cos(\theta_2+\alpha_5)v_2^2v_3^2\nonumber\\
&+\csc(\theta_1-\theta_2-\alpha_3)\sin(\theta_1-\alpha_3+\alpha_5)v_3^4\big)\Biggl), \label{charged mass 16}
\end{align}
\begin{align}
M^{2}_{\pm23}=&\frac{1}{2v_{23}^2v}\Biggl(4v_1v_2v_3v^2\csc(\theta_1-\theta_2-\alpha_3)\bigl(\lambda_2\cos\alpha_3\sin(\theta_1-\theta_2)\nonumber\\
&+\lambda_3\sin\alpha_3\cos(\theta_1-\theta_2)\bigl)+\lambda_4\Bigl(-\bigl(\cos(\theta_2+\beta_4)-i\sin(\theta_2+\beta_4)\bigl)v_1^3v_2^2\nonumber\\
&+\csc(\theta_1+\alpha_4)\bigl(2\sin(\theta_1-\theta_2+\alpha_4-\beta_4)-\sin(\theta_1-\theta_2-\alpha_4+\beta_4)\bigl)v_1^3v_2^2\nonumber\\
&-\csc(\theta_1+\alpha_4)\sin(\theta_1-\theta_2+\alpha_4-\beta_4)v_1v_2^4+\bigl(\cos\theta_1-i\sin\theta_1\bigl)\nonumber\\
&\times\bigl(3\cos(\theta_1-\theta_2)+i\sin(\theta_1-\theta_2)\bigl)\bigl(\cos\beta_4-i\sin\beta_4\bigl)v_1v_2^4\nonumber\\
&+\bigl(\cos(\theta_2+\beta_4)+i\sin(\theta_2+\beta_4)\bigl)v_1^3v_3^2+\csc(\theta_1-\theta_2-\alpha_3)\nonumber\\
&\times\bigl(-2\sin(\theta_1-\alpha_3+\beta_4)+\sin(\theta_1+\alpha_3+\beta_4)\bigl)v_1^3v_3^2\nonumber\\
&-3\bigl(\cos\theta_1-i\sin\theta_1\bigl)\bigl(\cos(\theta_1+\theta_2+\beta_4)+i\sin(\theta_1+\theta_2+\beta_4)\bigl)v_1v_2^2v_3^2\nonumber\\
&+\csc(\theta_1-\theta_2-\alpha_3)\bigl(\sin(\theta_1-2\theta_2-\alpha_3-\beta_4)+2\sin(\theta_1-\alpha_3+\beta_4)\bigl)v_1v_2^2v_3^2\nonumber\\
&-2\cos(\theta_1+\beta_4)\bigl(\cos(\theta_1-\theta_2)-i\sin(\theta_1-\theta_2)\bigl)v_1v_3^4\nonumber\\
&-\bigl(-i+\cot(\theta_1+\alpha_4)\bigl)\bigl(\sin(2\theta_1-\theta_2+\beta_4)+2\sin(\theta_2+\beta_4)\bigl)v_1v_2^2v_{23}^2\nonumber\\
&+\frac{v_2^4v_{23}^2}{v_1}\bigl(-i+\cot(\theta_1+\alpha_4)\bigl)\sin(\theta_2+\beta_4)+\bigl(-i+\cot(\theta_1-\theta_2-\alpha_3)\bigl)\nonumber\\
&\times\bigl(\sin(2\theta_1-\theta_2+\beta_4)-2\sin(\theta_2+\beta_4)\bigl)v_1v_3^2v_{23}^2\nonumber\\
&+\frac{v_2^2v_3^2v_{23}^2}{v_1}\bigl(-i+\cot(\theta_1-\theta_2-\alpha_3)\bigl)\sin(\theta_2+\beta_4)\Bigl)\nonumber\\
&+\lambda_6v_1v_2v_3v^2-\lambda_7\frac{2v_2v_3v^2}{v_1}\biggl(\csc(\theta_1+\alpha_4)\sin(\theta_1-\alpha_4+\beta_7)v_1^2\nonumber\\
&+\bigl(\cos\theta_1-i\sin\theta_1\bigl)\sin(2\theta_2+\beta_7)\Bigl(\csc(\theta_1+\alpha_4)\bigl(\cos\alpha_4-i\sin\alpha_4\bigl)v_2^2\nonumber\\
&+\csc(\theta_1-\theta_2-\alpha_3)\bigl(\cos(\theta_2+\alpha_3)+i\sin(\theta_2+\alpha_3)\bigl)v_3^2\Bigl)\biggl)\nonumber\\
&-\mu_5^2\frac{v^2}{v_1}\bigl(\cos\theta_1-i\sin\theta_1\bigl)\sin(\theta_2+\alpha_5)\Bigl(\csc(\theta_1+\alpha_4)\bigl(\cos\alpha_4-i\sin\alpha_4\bigl)v_2^2\nonumber\\
&+\csc(\theta_1-\theta_2-\alpha_3)\bigl(\cos(\theta_2+\alpha_3)+i\sin(\theta_2+\alpha_3)\bigl)v_3^2\Bigl)\Biggl),\label{charged mass}
\end{align}
\begin{align}
M^{2}_{\pm33}=&\frac{1}{v_{23}^2}\Biggl(2v_2^2v^2\csc(\theta_1-\theta_2-\alpha_3)\bigl(\lambda_2\cos\alpha_3\sin(\theta_1-\theta_2)+\lambda_3\sin\alpha_3\cos(\theta_1-\theta_2)\bigl)\nonumber\\
&+\lambda_4\frac{v_2v_3v^2}{4v_1^2}\csc(\theta_1-\theta_2-\alpha_3)\csc(\theta_1+\alpha_4)\Bigl(\bigl(\cos(2\theta_1-2\theta_2-\alpha_3+\alpha_4-\beta_4)\nonumber\\
&-3\cos(\alpha_3+\alpha_4-\beta_4)-\cos(2\theta_1-2\theta_2-\alpha_3-\alpha_4+\beta_4)\nonumber\\
&+2\cos(\alpha_3-\alpha_4+\beta_4)+\cos(2\theta_1-\alpha_3+\alpha_4+\beta_4)\nonumber\\
&-\cos(2\theta_1+\alpha_3+\alpha_4+\beta_4)+\cos(2\theta_2+\alpha_3+\alpha_4+\beta_4)\bigl)v_1^2\nonumber\\
&+2\sin(\theta_2+\alpha_3+\alpha_4)\sin(\theta_2+\beta_4)v_2^2\Bigl)-\frac{1}{2}\lambda_6v_3^2v^2\nonumber\\
&+\lambda_7\frac{v_3^2v^2}{v_1^2}\csc(\theta_1+\alpha_4)\bigl(\sin(\theta_1-\alpha_4+\beta_7)v_1^2\nonumber\\
&-\csc(\theta_1-\theta_2-\alpha_3)\sin(\theta_2+\alpha_3+\alpha_4)\sin(2\theta_2+\beta_7)v_2^2\bigl)\nonumber\\
&-\mu_5^2\frac{v_2v_3v^2}{2v_1^2}\csc(\theta_1-\theta_2-\alpha_3)\csc(\theta_1+\alpha_4)\sin(\theta_2+\alpha_3+\alpha_4)\sin(\theta_2+\alpha_5)\Biggl).  \label{charged mass 18}
\end{align}

\section{Yukawa Couplings}

The numerical Yukawa coupling matrices~\autoref{Yukawa_matrices } in the $S_3$ basis, corresponding to the best fits for quark and lepton masses and mixings, are presented here:
\begin{align}
K_{d}^1 &= \begin{pmatrix}
    0 & 1.50\times10^{-4}+9.86\times10^{-5}i & -1.57\times10^{-4}+5.71\times10^{-5}i\\[6pt]
     1.50\times10^{-4}+9.86\times10^{-5}i & 0 & 0 \\[6pt]
    5.11\times10^{-3}-4.57\times10^{-3}i & 0 & 0\\[6pt]
\end{pmatrix},\label{C1}
\end{align}
\begin{align}
K_{d}^2 &= \begin{pmatrix}
    1.50\times10^{-4}+9.86\times10^{-5}i  & 0 & 0 \\[6pt]
    0 & -1.50\times10^{-4}-9.86\times10^{-5}i  & -1.57\times10^{-4}+5.71\times10^{-5}i \\[6pt]
    0 & 5.11\times10^{-3}-4.57\times10^{-3}i & 0\\[6pt]
\end{pmatrix},\label{C2}
\end{align}
\begin{align}
K_{d}^3 &= \begin{pmatrix}
    -1.82\times10^{-4}-2.81\times10^{-4}i & 0 & 0 \\[6pt]
    0 & -1.82\times10^{-4}-2.81\times10^{-4}i& 0 \\[6pt]
    0 & 0 & -4.15\times10^{-4}-1.20\times10^{-2}i\\[6pt]
\end{pmatrix}, \label{C3}
\end{align}
\begin{align}
K_{u}^1 &= \begin{pmatrix}
    0 & 1.57\times10^{-7}-6.02\times10^{-6}i &1.41\times10^{-2}+6.86\times10^{-3}i\\[6pt]
    1.57\times10^{-7}-6.02\times10^{-6}i & 0 & 0 \\[6pt]
    6.03\times10^{-2}+1.51\times10^{-1}i & 0 & 0\\[6pt]
\end{pmatrix},\label{C4}
\end{align}
\begin{align}
K_{u}^2 &= \begin{pmatrix}
    1.57\times10^{-7}-6.02\times10^{-6}i& 0 & 0 \\[6pt]
    0 & -1.57\times10^{-7}+6.02\times10^{-6}i& 1.41\times10^{-2}+6.86\times10^{-3}i \\[6pt]
    0 & 6.03\times10^{-2}+1.51\times10^{-1}i & 0\\[6pt]
\end{pmatrix},\label{C5}
\end{align}
\begin{align}
K_{u}^3 &= \begin{pmatrix}
    -2.97\times10^{-6}+1.85\times10^{-7}i & 0 & 0 \\[6pt]
    0 & -2.97\times10^{-6}+1.85\times10^{-7}i & 0 \\[6pt]
    0 & 0 & 1.41\times10^{-2}+6.86\times10^{-3}i\\[6pt]
\end{pmatrix}.\label{C6}
\end{align}

\bibliographystyle{bibsty}
\bibliography{references}

\providecommand{\href}[2]{#2}\begingroup\raggedright\begin{thebibliography}{10}

\bibitem{ATLAS:2012yve}
{\bf ATLAS} Collaboration, G.~Aad et~al., {\it {Observation of a new particle
  in the search for the Standard Model Higgs boson with the ATLAS detector at
  the LHC}},  {\em Phys. Lett. B} {\bf 716} (2012) 1--29,
  [\href{http://arxiv.org/abs/1207.7214}{{\tt arXiv:1207.7214}}].

\bibitem{CMS:2012qbp}
{\bf CMS} Collaboration, S.~Chatrchyan et~al., {\it {Observation of a New Boson
  at a Mass of 125 GeV with the CMS Experiment at the LHC}},  {\em Phys. Lett.
  B} {\bf 716} (2012) 30--61, [\href{http://arxiv.org/abs/1207.7235}{{\tt
  arXiv:1207.7235}}].

\bibitem{Profumo:2007wc}
S.~Profumo, M.~J. Ramsey-Musolf, and G.~Shaughnessy, {\it {Singlet Higgs
  phenomenology and the electroweak phase transition}},  {\em JHEP} {\bf 08}
  (2007) 010, [\href{http://arxiv.org/abs/0705.2425}{{\tt arXiv:0705.2425}}].

\bibitem{Chiang:2017nmu}
C.-W. Chiang, M.~J. Ramsey-Musolf, and E.~Senaha, {\it {Standard Model with a
  Complex Scalar Singlet: Cosmological Implications and Theoretical
  Considerations}},  {\em Phys. Rev. D} {\bf 97} (2018), no.~1 015005,
  [\href{http://arxiv.org/abs/1707.09960}{{\tt arXiv:1707.09960}}].

\bibitem{Branco:2011iw}
G.~C. Branco, P.~M. Ferreira, L.~Lavoura, M.~N. Rebelo, M.~Sher, and J.~P.
  Silva, {\it {Theory and phenomenology of two-Higgs-doublet models}},  {\em
  Phys. Rept.} {\bf 516} (2012) 1--102,
  [\href{http://arxiv.org/abs/1106.0034}{{\tt arXiv:1106.0034}}].

\bibitem{Fontes:2017zfn}
D.~Fontes, M.~M\"uhlleitner, J.~C. Rom\~ao, R.~Santos, J.~a.~P. Silva, and
  J.~Wittbrodt, {\it {The C2HDM revisited}},  {\em JHEP} {\bf 02} (2018) 073,
  [\href{http://arxiv.org/abs/1711.09419}{{\tt arXiv:1711.09419}}].

\bibitem{Georgi:1985nv}
H.~Georgi and M.~Machacek, {\it {DOUBLY CHARGED HIGGS BOSONS}},  {\em Nucl.
  Phys. B} {\bf 262} (1985) 463--477.

\bibitem{Hartling:2014zca}
K.~Hartling, K.~Kumar, and H.~E. Logan, {\it {The decoupling limit in the
  Georgi-Machacek model}},  {\em Phys. Rev. D} {\bf 90} (2014), no.~1 015007,
  [\href{http://arxiv.org/abs/1404.2640}{{\tt arXiv:1404.2640}}].

\bibitem{Pakvasa:1977in}
S.~Pakvasa and H.~Sugawara, {\it {Discrete Symmetry and Cabibbo Angle}},  {\em
  Phys. Lett. B} {\bf 73} (1978) 61--64.

\bibitem{Derman:1978rx}
E.~Derman, {\it {Flavor Unification, $\tau$ Decay and $b$ Decay Within the Six
  Quark Six Lepton {Weinberg-Salam} Model}},  {\em Phys. Rev. D} {\bf 19}
  (1979) 317--329.

\bibitem{Derman:1979nf}
E.~Derman and H.-S. Tsao, {\it {SU(2) X U(1) X S($n$) Flavor Dynamics and a
  Bound on the Number of Flavors}},  {\em Phys. Rev. D} {\bf 20} (1979) 1207.

\bibitem{Wyler:1978fj}
D.~Wyler, {\it {The Cabibbo Angle in the SU(2)$_L \times$ U(1) Gauge
  Theories}},  {\em Phys. Rev. D} {\bf 19} (1979) 330.

\bibitem{Frere:1978ds}
J.~M. Frere, {\it {On the Use of Permutation Symmetry}},  {\em Phys. Lett. B}
  {\bf 80} (1979) 369--371.

\bibitem{Yahalom:1983kf}
R.~Yahalom, {\it {Horizontal Permutation Symmetry, Fermion Masses and
  Pseudogoldstone Bosons in SU(2)$_L \times$ U(1)}},  {\em Phys. Rev. D} {\bf
  29} (1984) 536.

\bibitem{Ma:1990qh}
E.~Ma, {\it {Two derivable relationships among quark masses and mixing
  angles}},  {\em Phys. Rev. D} {\bf 43} (1991) 2761--2764.

\bibitem{Hall:1995es}
L.~J. Hall and H.~Murayama, {\it {A Geometry of the generations}},  {\em Phys.
  Rev. Lett.} {\bf 75} (1995) 3985--3988,
  [\href{http://arxiv.org/abs/hep-ph/9508296}{{\tt hep-ph/9508296}}].

\bibitem{Koide:1999mx}
Y.~Koide, {\it {Universal seesaw mass matrix model with an S(3) symmetry}},
  {\em Phys. Rev. D} {\bf 60} (1999) 077301,
  [\href{http://arxiv.org/abs/hep-ph/9905416}{{\tt hep-ph/9905416}}].

\bibitem{Lavoura:1999dn}
L.~Lavoura, {\it {A New model for the quark mass matrices}},  {\em Phys. Rev.
  D} {\bf 61} (2000) 077303, [\href{http://arxiv.org/abs/hep-ph/9907538}{{\tt
  hep-ph/9907538}}].

\bibitem{Kubo:2003iw}
J.~Kubo, A.~Mondragon, M.~Mondragon, and E.~Rodriguez-Jauregui, {\it {The
  Flavor symmetry}},  {\em Prog. Theor. Phys.} {\bf 109} (2003) 795--807,
  [\href{http://arxiv.org/abs/hep-ph/0302196}{{\tt hep-ph/0302196}}]. [Erratum:
  Prog.Theor.Phys. 114, 287--287 (2005)].

\bibitem{Kubo:2004ps}
J.~Kubo, H.~Okada, and F.~Sakamaki, {\it {Higgs potential in minimal S(3)
  invariant extension of the standard model}},  {\em Phys. Rev. D} {\bf 70}
  (2004) 036007, [\href{http://arxiv.org/abs/hep-ph/0402089}{{\tt
  hep-ph/0402089}}].

\bibitem{Chen:2004rr}
S.-L. Chen, M.~Frigerio, and E.~Ma, {\it {Large neutrino mixing and normal mass
  hierarchy: A Discrete understanding}},  {\em Phys. Rev. D} {\bf 70} (2004)
  073008, [\href{http://arxiv.org/abs/hep-ph/0404084}{{\tt hep-ph/0404084}}].
  [Erratum: Phys.Rev.D 70, 079905 (2004)].

\bibitem{Kimura:2005sx}
T.~Kimura, {\it {The minimal S(3) symmetric model}},  {\em Prog. Theor. Phys.}
  {\bf 114} (2005) 329--358.

\bibitem{Teshima:2005bk}
T.~Teshima, {\it {Flavor mass and mixing and S(3) symmetry: An S(3) invariant
  model reasonable to all}},  {\em Phys. Rev. D} {\bf 73} (2006) 045019,
  [\href{http://arxiv.org/abs/hep-ph/0509094}{{\tt hep-ph/0509094}}].

\bibitem{Koide:2005ep}
Y.~Koide, {\it {Permutation symmetry S(3) and VEV structure of flavor-triplet
  Higgs scalars}},  {\em Phys. Rev. D} {\bf 73} (2006) 057901,
  [\href{http://arxiv.org/abs/hep-ph/0509214}{{\tt hep-ph/0509214}}].

\bibitem{Araki:2005ec}
T.~Araki, J.~Kubo, and E.~A. Paschos, {\it {S(3) flavor symmetry and
  leptogenesis}},  {\em Eur. Phys. J. C} {\bf 45} (2006) 465--475,
  [\href{http://arxiv.org/abs/hep-ph/0502164}{{\tt hep-ph/0502164}}].

\bibitem{Mondragon:2007nk}
A.~Mondragon, M.~Mondragon, and E.~Peinado, {\it {S(3)-flavour symmetry as
  realized in lepton flavour violating processes}},  {\em J. Phys. A} {\bf 41}
  (2008) 304035, [\href{http://arxiv.org/abs/0712.1799}{{\tt
  arXiv:0712.1799}}].

\bibitem{Mondragon:2007af}
A.~Mondragon, M.~Mondragon, and E.~Peinado, {\it {Lepton masses, mixings and
  FCNC in a minimal S(3)-invariant extension of the Standard Model}},  {\em
  Phys. Rev. D} {\bf 76} (2007) 076003,
  [\href{http://arxiv.org/abs/0706.0354}{{\tt arXiv:0706.0354}}].

\bibitem{Bhattacharyya:2010hp}
G.~Bhattacharyya, P.~Leser, and H.~Pas, {\it {Exotic Higgs boson decay modes as
  a harbinger of $S_3$ flavor symmetry}},  {\em Phys. Rev. D} {\bf 83} (2011)
  011701, [\href{http://arxiv.org/abs/1006.5597}{{\tt arXiv:1006.5597}}].

\bibitem{Teshima:2011wg}
T.~Teshima and Y.~Okumura, {\it {Quark/lepton mass and mixing in $S_3$
  invariant model and CP-violation of neutrino}},  {\em Phys. Rev. D} {\bf 84}
  (2011) 016003, [\href{http://arxiv.org/abs/1103.6127}{{\tt
  arXiv:1103.6127}}].

\bibitem{Teshima:2012cg}
T.~Teshima, {\it {Higgs potential in $S_3$ invariant model for quark/lepton
  mass and mixing}},  {\em Phys. Rev. D} {\bf 85} (2012) 105013,
  [\href{http://arxiv.org/abs/1202.4528}{{\tt arXiv:1202.4528}}].

\bibitem{Bhattacharyya:2012ze}
G.~Bhattacharyya, P.~Leser, and H.~Pas, {\it {Novel signatures of the Higgs
  sector from S3 flavor symmetry}},  {\em Phys. Rev. D} {\bf 86} (2012) 036009,
  [\href{http://arxiv.org/abs/1206.4202}{{\tt arXiv:1206.4202}}].

\bibitem{GonzalezCanales:2012blg}
F.~Gonzalez~Canales, A.~Mondragon, and M.~Mondragon, {\it {The $S_3$ Flavour
  Symmetry: Neutrino Masses and Mixings}},  {\em Fortsch. Phys.} {\bf 61}
  (2013) 546--570, [\href{http://arxiv.org/abs/1205.4755}{{\tt
  arXiv:1205.4755}}].

\bibitem{GonzalezCanales:2013pdx}
F.~Gonz\'alez~Canales, A.~Mondrag\'on, M.~Mondrag\'on, U.~J. Salda\~na Salazar,
  and L.~Velasco-Sevilla, {\it {Quark sector of S3 models: classification and
  comparison with experimental data}},  {\em Phys. Rev. D} {\bf 88} (2013)
  096004, [\href{http://arxiv.org/abs/1304.6644}{{\tt arXiv:1304.6644}}].

\bibitem{Das:2014fea}
D.~Das and U.~K. Dey, {\it {Analysis of an extended scalar sector with $S_3$
  symmetry}},  {\em Phys. Rev. D} {\bf 89} (2014), no.~9 095025,
  [\href{http://arxiv.org/abs/1404.2491}{{\tt arXiv:1404.2491}}]. [Erratum:
  Phys.Rev.D 91, 039905 (2015)].

\bibitem{Das:2015sca}
D.~Das, U.~K. Dey, and P.~B. Pal, {\it {$S_3$ symmetry and the quark mixing
  matrix}},  {\em Phys. Lett. B} {\bf 753} (2016) 315--318,
  [\href{http://arxiv.org/abs/1507.06509}{{\tt arXiv:1507.06509}}].

\bibitem{Emmanuel-Costa:2016vej}
D.~Emmanuel-Costa, O.~M. Ogreid, P.~Osland, and M.~N. Rebelo, {\it {Spontaneous
  symmetry breaking in the $S_3$-symmetric scalar sector}},  {\em JHEP} {\bf
  02} (2016) 154, [\href{http://arxiv.org/abs/1601.04654}{{\tt
  arXiv:1601.04654}}]. [Erratum: JHEP 08, 169 (2016)].

\bibitem{Kuncinas:2020wrn}
A.~Kun\v{c}inas, O.~M. Ogreid, P.~Osland, and M.~N. Rebelo, {\it {S3 -inspired
  three-Higgs-doublet models: A class with a complex vacuum}},  {\em Phys. Rev.
  D} {\bf 101} (2020), no.~7 075052,
  [\href{http://arxiv.org/abs/2001.01994}{{\tt arXiv:2001.01994}}].

\bibitem{Gomez-Bock:2021uyu}
M.~G\'omez-Bock, M.~Mondrag\'on, and A.~P\'erez-Mart\'\i{}nez, {\it {Scalar and
  gauge sectors in the 3-Higgs Doublet Model under the $S_3$ symmetry}},  {\em
  Eur. Phys. J. C} {\bf 81} (2021), no.~10 942,
  [\href{http://arxiv.org/abs/2102.02800}{{\tt arXiv:2102.02800}}].

\bibitem{Khater:2021wcx}
W.~Khater, A.~Kun\v{c}inas, O.~M. Ogreid, P.~Osland, and M.~N. Rebelo, {\it
  {Dark matter in three-Higgs-doublet models with S$_{3}$ symmetry}},  {\em
  JHEP} {\bf 01} (2022) 120, [\href{http://arxiv.org/abs/2108.07026}{{\tt
  arXiv:2108.07026}}].

\bibitem{Kuncinas:2022whn}
A.~Kun\v{c}inas, O.~M. Ogreid, P.~Osland, and M.~N. Rebelo, {\it {Dark matter
  in a CP-violating three-Higgs-doublet model with S3 symmetry}},  {\em Phys.
  Rev. D} {\bf 106} (2022), no.~7 075002,
  [\href{http://arxiv.org/abs/2204.05684}{{\tt arXiv:2204.05684}}].

\bibitem{Yamanaka:1981pa}
Y.~Yamanaka, H.~Sugawara, and S.~Pakvasa, {\it {Permutation Symmetries and the
  Fermion Mass Matrix}},  {\em Phys. Rev. D} {\bf 25} (1982) 1895. [Erratum:
  Phys.Rev.D 29, 2135 (1984)].

\bibitem{Brown:1984mq}
T.~Brown, N.~Deshpande, S.~Pakvasa, and H.~Sugawara, {\it {{CP} Nonconservation
  and Rare Processes in S(4) Model of Permutation Symmetry}},  {\em Phys. Lett.
  B} {\bf 141} (1984) 95--99.

\bibitem{CarcamoHernandez:2022vjk}
A.~E. C\'arcamo~Hern\'andez, C.~Espinoza, J.~C. G\'omez-Izquierdo, J.~M.
  Gonz\'alez, and M.~Mondrag\'on, {\it {Predictive extended 3HDM with $S_4$
  family symmetry}},  \href{http://arxiv.org/abs/2212.12000}{{\tt
  arXiv:2212.12000}}.

\bibitem{Pramanick:2017wry}
S.~Pramanick and A.~Raychaudhuri, {\it {Three-Higgs-doublet model under A4
  symmetry implies alignment}},  {\em JHEP} {\bf 01} (2018) 011,
  [\href{http://arxiv.org/abs/1710.04433}{{\tt arXiv:1710.04433}}].

\bibitem{Buskin:2021eig}
N.~Buskin and I.~P. Ivanov, {\it {Bounded-from-below conditions for
  $A_4$-symmetric 3HDM}},  {\em J. Phys. A} {\bf 54} (2021) 325401,
  [\href{http://arxiv.org/abs/2104.11428}{{\tt arXiv:2104.11428}}].

\bibitem{Carrolo:2022oyg}
S.~Carrolo, J.~C. Romao, and J.~P. Silva, {\it {Conditions for global minimum
  in the $A_4$ symmetric 3HDM}},  {\em Eur. Phys. J. C} {\bf 82} (2022), no.~8
  749, [\href{http://arxiv.org/abs/2207.02928}{{\tt arXiv:2207.02928}}].

\bibitem{Ivanov:2012fp}
I.~P. Ivanov and E.~Vdovin, {\it {Classification of finite reparametrization
  symmetry groups in the three-Higgs-doublet model}},  {\em Eur. Phys. J. C}
  {\bf 73} (2013), no.~2 2309, [\href{http://arxiv.org/abs/1210.6553}{{\tt
  arXiv:1210.6553}}].

\bibitem{Emmanuel-Costa:2017zkm}
D.~Emmanuel-Costa, O.~M. Ogreid, P.~Osland, and M.~N. Rebelo, {\it {Spontaneous
  symmetry breaking in three-Higgs-doublet $S_3$-symmetric models}},  {\em J.
  Phys. Conf. Ser.} {\bf 873} (2017), no.~1 012007,
  [\href{http://arxiv.org/abs/1703.08457}{{\tt arXiv:1703.08457}}].

\bibitem{Minkowski:1977sc}
P.~Minkowski, {\it {$\mu \to e\gamma$ at a Rate of One Out of $10^{9}$ Muon
  Decays?}},  {\em Phys. Lett. B} {\bf 67} (1977) 421--428.

\bibitem{Yanagida:1979as}
T.~Yanagida, {\it {Horizontal gauge symmetry and masses of neutrinos}},  {\em
  Conf. Proc. C} {\bf 7902131} (1979) 95--99.

\bibitem{Mohapatra:1979ia}
R.~N. Mohapatra and G.~Senjanovic, {\it {Neutrino Mass and Spontaneous Parity
  Nonconservation}},  {\em Phys. Rev. Lett.} {\bf 44} (1980) 912.

\bibitem{Glashow:1979nm}
S.~L. Glashow, {\it {The Future of Elementary Particle Physics}},  {\em NATO
  Sci. Ser. B} {\bf 61} (1980) 687.

\bibitem{Faro:2019vcd}
F.~S. Faro and I.~P. Ivanov, {\it {Boundedness from below in the $U(1)\times
  U(1)$ three-Higgs-doublet model}},  {\em Phys. Rev. D} {\bf 100} (2019),
  no.~3 035038, [\href{http://arxiv.org/abs/1907.01963}{{\tt
  arXiv:1907.01963}}].

\bibitem{Deshpande:1994en}
N.~G. Deshpande and X.-G. He, {\it {CP violation in a multi - Higgs doublet
  model}},  {\em Pramana} {\bf 45} (1995) S73--S83,
  [\href{http://arxiv.org/abs/hep-ph/9409234}{{\tt hep-ph/9409234}}].

\bibitem{Babu:2018uik}
K.~S. Babu and S.~Jana, {\it {Enhanced Di-Higgs Production in the Two Higgs
  Doublet Model}},  {\em JHEP} {\bf 02} (2019) 193,
  [\href{http://arxiv.org/abs/1812.11943}{{\tt arXiv:1812.11943}}].

\bibitem{Babu:2009nn}
K.~S. Babu and Y.~Meng, {\it {Flavor Violation in Supersymmetric Q(6) Model}},
  {\em Phys. Rev. D} {\bf 80} (2009) 075003,
  [\href{http://arxiv.org/abs/0907.4231}{{\tt arXiv:0907.4231}}].

\bibitem{Ciuchini:1998ix}
M.~Ciuchini et~al., {\it {Delta M(K) and epsilon(K) in SUSY at the
  next-to-leading order}},  {\em JHEP} {\bf 10} (1998) 008,
  [\href{http://arxiv.org/abs/hep-ph/9808328}{{\tt hep-ph/9808328}}].

\bibitem{Becirevic:2001jj}
D.~Becirevic, M.~Ciuchini, E.~Franco, V.~Gimenez, G.~Martinelli, A.~Masiero,
  M.~Papinutto, J.~Reyes, and L.~Silvestrini, {\it {$B_d - \bar{B}_d$ mixing
  and the $B_d \to J/\psi K_s$ asymmetry in general SUSY models}},  {\em Nucl.
  Phys. B} {\bf 634} (2002) 105--119,
  [\href{http://arxiv.org/abs/hep-ph/0112303}{{\tt hep-ph/0112303}}].

\bibitem{UTfit:2007eik}
{\bf UTfit} Collaboration, M.~Bona et~al., {\it {Model-independent constraints
  on $\Delta F=2$ operators and the scale of new physics}},  {\em JHEP} {\bf
  03} (2008) 049, [\href{http://arxiv.org/abs/0707.0636}{{\tt
  arXiv:0707.0636}}].

\bibitem{Bauer:2009cf}
M.~Bauer, S.~Casagrande, U.~Haisch, and M.~Neubert, {\it {Flavor Physics in the
  Randall-Sundrum Model: II. Tree-Level Weak-Interaction Processes}},  {\em
  JHEP} {\bf 09} (2010) 017, [\href{http://arxiv.org/abs/0912.1625}{{\tt
  arXiv:0912.1625}}].

\bibitem{Abel:2020pzs}
C.~Abel et~al., {\it {Measurement of the Permanent Electric Dipole Moment of
  the Neutron}},  {\em Phys. Rev. Lett.} {\bf 124} (2020), no.~8 081803,
  [\href{http://arxiv.org/abs/2001.11966}{{\tt arXiv:2001.11966}}].

\bibitem{Ibrahim:2007fb}
T.~Ibrahim and P.~Nath, {\it {CP Violation From Standard Model to Strings}},
  {\em Rev. Mod. Phys.} {\bf 80} (2008) 577--631,
  [\href{http://arxiv.org/abs/0705.2008}{{\tt arXiv:0705.2008}}].

\bibitem{Iltan:2001vg}
E.~O. Iltan, {\it {Top quark electric and chromo electric dipole moments in the
  general two Higgs doublet model}},  {\em Phys. Rev. D} {\bf 65} (2002)
  073013, [\href{http://arxiv.org/abs/hep-ph/0111038}{{\tt hep-ph/0111038}}].

\bibitem{Bertolini:2019out}
S.~Bertolini, A.~Maiezza, and F.~Nesti, {\it {Kaon CP violation and neutron EDM
  in the minimal left-right symmetric model}},  {\em Phys. Rev. D} {\bf 101}
  (2020), no.~3 035036, [\href{http://arxiv.org/abs/1911.09472}{{\tt
  arXiv:1911.09472}}].

\bibitem{Babu:2009fd}
K.~S. Babu, {\it {TASI Lectures on Flavor Physics}},  in {\em {Theoretical
  Advanced Study Institute in Elementary Particle Physics}: {The Dawn of the
  LHC Era}}, pp.~49--123, 2010.
\newblock \href{http://arxiv.org/abs/0910.2948}{{\tt arXiv:0910.2948}}.

\bibitem{ParticleDataGroup:2020ssz}
{\bf Particle Data Group} Collaboration, P.~A. Zyla et~al., {\it {Review of
  Particle Physics}},  {\em PTEP} {\bf 2020} (2020), no.~8 083C01.

\bibitem{Feroz:2008xx}
F.~Feroz, M.~P. Hobson, and M.~Bridges, {\it {MultiNest: an efficient and
  robust Bayesian inference tool for cosmology and particle physics}},  {\em
  Mon. Not. Roy. Astron. Soc.} {\bf 398} (2009) 1601--1614,
  [\href{http://arxiv.org/abs/0809.3437}{{\tt arXiv:0809.3437}}].

\bibitem{Feroz:2013hea}
F.~Feroz, M.~P. Hobson, E.~Cameron, and A.~N. Pettitt, {\it {Importance Nested
  Sampling and the MultiNest Algorithm}},  {\em Open J. Astrophys.} {\bf 2}
  (2019), no.~1 10, [\href{http://arxiv.org/abs/1306.2144}{{\tt
  arXiv:1306.2144}}].

\bibitem{Buchner:2014nha}
J.~Buchner, A.~Georgakakis, K.~Nandra, L.~Hsu, C.~Rangel, M.~Brightman,
  A.~Merloni, M.~Salvato, J.~Donley, and D.~Kocevski, {\it {X-ray spectral
  modelling of the AGN obscuring region in the CDFS: Bayesian model selection
  and catalogue}},  {\em Astron. Astrophys.} {\bf 564} (2014) A125,
  [\href{http://arxiv.org/abs/1402.0004}{{\tt arXiv:1402.0004}}].

\bibitem{Esteban:2020cvm}
I.~Esteban, M.~C. Gonzalez-Garcia, M.~Maltoni, T.~Schwetz, and A.~Zhou, {\it
  {The fate of hints: updated global analysis of three-flavor neutrino
  oscillations}},  {\em JHEP} {\bf 09} (2020) 178,
  [\href{http://arxiv.org/abs/2007.14792}{{\tt arXiv:2007.14792}}].

\bibitem{Huang:2019tdh}
G.-y. Huang, W.~Rodejohann, and S.~Zhou, {\it {Effective neutrino masses in
  KATRIN and future tritium beta-decay experiments}},  {\em Phys. Rev. D} {\bf
  101} (2020), no.~1 016003, [\href{http://arxiv.org/abs/1910.08332}{{\tt
  arXiv:1910.08332}}].

\bibitem{Bilenky:2014uka}
S.~M. Bilenky and C.~Giunti, {\it {Neutrinoless Double-Beta Decay: a Probe of
  Physics Beyond the Standard Model}},  {\em Int. J. Mod. Phys. A} {\bf 30}
  (2015), no.~04n05 1530001, [\href{http://arxiv.org/abs/1411.4791}{{\tt
  arXiv:1411.4791}}].

\bibitem{RoyChoudhury:2018gay}
S.~Roy~Choudhury and S.~Choubey, {\it {Updated Bounds on Sum of Neutrino Masses
  in Various Cosmological Scenarios}},  {\em JCAP} {\bf 09} (2018) 017,
  [\href{http://arxiv.org/abs/1806.10832}{{\tt arXiv:1806.10832}}].

\bibitem{Formaggio:2021nfz}
J.~A. Formaggio, A.~L.~C. de~Gouv\^ea, and R.~G.~H. Robertson, {\it {Direct
  Measurements of Neutrino Mass}},  {\em Phys. Rept.} {\bf 914} (2021) 1--54,
  [\href{http://arxiv.org/abs/2102.00594}{{\tt arXiv:2102.00594}}].

\bibitem{Fogli:2004as}
G.~L. Fogli, E.~Lisi, A.~Marrone, A.~Melchiorri, A.~Palazzo, P.~Serra, and
  J.~Silk, {\it {Observables sensitive to absolute neutrino masses: Constraints
  and correlations from world neutrino data}},  {\em Phys. Rev. D} {\bf 70}
  (2004) 113003, [\href{http://arxiv.org/abs/hep-ph/0408045}{{\tt
  hep-ph/0408045}}].

\bibitem{ParticleDataGroup:2022pth}
{\bf Particle Data Group} Collaboration, R.~L. Workman et~al., {\it {Review of
  Particle Physics}},  {\em PTEP} {\bf 2022} (2022) 083C01.

\bibitem{KamLAND-Zen:2016pfg}
{\bf KamLAND-Zen} Collaboration, A.~Gando et~al., {\it {Search for Majorana
  Neutrinos near the Inverted Mass Hierarchy Region with KamLAND-Zen}},  {\em
  Phys. Rev. Lett.} {\bf 117} (2016), no.~8 082503,
  [\href{http://arxiv.org/abs/1605.02889}{{\tt arXiv:1605.02889}}]. [Addendum:
  Phys.Rev.Lett. 117, 109903 (2016)].

\bibitem{Planck:2018vyg}
{\bf Planck} Collaboration, N.~Aghanim et~al., {\it {Planck 2018 results. VI.
  Cosmological parameters}},  {\em Astron. Astrophys.} {\bf 641} (2020) A6,
  [\href{http://arxiv.org/abs/1807.06209}{{\tt arXiv:1807.06209}}]. [Erratum:
  Astron.Astrophys. 652, C4 (2021)].

\bibitem{Cepeda:2019klc}
M.~Cepeda et~al., {\it {Report from Working Group 2}: {Higgs Physics at the
  HL-LHC and HE-LHC}},  {\em CERN Yellow Rep. Monogr.} {\bf 7} (2019) 221--584,
  [\href{http://arxiv.org/abs/1902.00134}{{\tt arXiv:1902.00134}}].

\bibitem{FCC:2018byv}
{\bf FCC} Collaboration, A.~Abada et~al., {\it {FCC Physics Opportunities}:
  {Future Circular Collider Conceptual Design Report Volume 1}},  {\em Eur.
  Phys. J. C} {\bf 79} (2019), no.~6 474.

\bibitem{Accettura:2023ked}
C.~Accettura et~al., {\it {Towards a Muon Collider}},
  \href{http://arxiv.org/abs/2303.08533}{{\tt arXiv:2303.08533}}.

\bibitem{Alarcon:2022ero}
R.~Alarcon et~al., {\it {Electric dipole moments and the search for new
  physics}},  in {\em {Snowmass 2021}}, 3, 2022.
\newblock \href{http://arxiv.org/abs/2203.08103}{{\tt arXiv:2203.08103}}.

\end{thebibliography}\endgroup

\end{document}